\documentclass{jfm}

\pdfoutput=1

\usepackage{graphicx}
\usepackage{amssymb}
\usepackage[usenames]{color}

\newcommand{\be}{\begin{equation}}
\newcommand{\ee}{\end{equation}}
\newcommand{\bea}{\begin{eqnarray}}
\newcommand{\eea}{\end{eqnarray}}
\newcommand{\bfig}{\begin{figure}}
\newcommand{\efig}{\end{figure}}
\newcommand{\bc}{\begin{center}}
\newcommand{\ec}{\end{center}}
\newcommand{\btab}{\begin{tabular}}
\newcommand{\etab}{\end{tabular}}
\newcommand{\dr}{\partial}

\let\oldepsilon\epsilon
\let\epsilon\varepsilon
\let\varepsilon\oldepsilon

\newcommand{\Fr}{{\mathcal{F}}}

\newcommand{\Rey}{{\mathcal{R}}}

\title[]{Bedforms in a turbulent stream.\\Part 2: Formation of ripples by primary linear instability and of dunes by non-linear pattern coarsening} 
\author[A. Fourri\`ere, P. Claudin and B. Andreotti]
{A\ls N\ls T\ls O\ls I\ls N\ls E \ns F\ls O\ls U\ls R\ls R\ls I\ls E\ls R\ls E ,  \ns P\ls H\ls I\ls L\ls I\ls P\ls P\ls E \ns C\ls L\ls A\ls U\ls D\ls I\ls N \and \ns B\ls R\ls U\ls N\ls O\ns A\ls N\ls D\ls R\ls E\ls O\ls T\ls T\ls I \ns} 
\affiliation{
Laboratoire de Physique et M\'ecanique des Milieux H\'et\'erog\`enes\\
PMMH UMR 7636 CNRS-ESPCI-P6-P7,\\
10 rue Vauquelin, 75231 Paris Cedex 05, France.}
\pubyear{2000}
\volume{???}
\pagerange{??--??}
\date{\today}
\begin{document}
\maketitle

\begin{abstract}
It is widely accepted that both ripples and dunes form in rivers by primary linear instability, the wavelength of the former scaling on the grain size, that of the latter being controled by the water depth. We revisit here this problem, using the computation of the turbulent flow over a wavy bottom performed in Part~1. A multi-scale description of the problem is proposed, in which the details of the different mechanisms controlling sediment transport are encoded into three quantities: the saturated flux, the saturation length and the threshold shear stress. Theses quantities are modelled in the case of erosion and momentum limited bed loads. This framework allows to give a clear picture of the instability in terms of dynamical mechanisms. The relation between the wavelength at which ripples form and the flux saturation length is quantitatively derived. This solve the discrepancy between measured initial wavelengths and predictions that do not take this lag between flow velocity and sediment transport into account. Inverting the problem, experimental data is used to determine the saturation length as a function of grain size and shear velocity. Finally, using the systematic expansion of the flow field with respect to the corrugation amplitude performed in part 1, we discuss the non-linear selection of ripple aspect ratio. Investingating the effect of a free surface on the linear instability, we show that the excitation of standing waves at the surface has a stabilising effect, independently of the details of the flow and sediment transport models. Consequently, the shape of the dispersion relation obtained from the linear stability analysis of a flat sand bed is such that dunes can not result from a primary linear instability. We present the results of field experiments performed in the natural sandy Leyre river, which evidence the formation of ripples by a linear instability and the formation of dunes by a non-linear pattern coarsening limited by the free surface. Finally, we show that mega-dunes form when the sand bed presents heterogeneities such as a wide distribution of grain sizes.
\end{abstract}

%________________________________________________________________________
\section{Introduction}

Since \cite{R80}, and followed by others (\cite{SB84,McL90}), ripples and dunes observed on the bed of sandy rivers have been interpreted as the two most unstable modes of the same linear instability. Although the classification of river bedform types is a difficult task (\cite{A90}), subaqueous ripples and dunes are standardly defined by their typical size: ripples are the small scale bedforms whose wavelength $\lambda$ scales on the grain size and dunes are those whose wavelength is comparable to or larger than the flow depth $H$ (\cite{K63,R80,E70,F74,EF82,A85}). A variant of this definition was used by \cite{GSR66}: dunes are the bed features larger than ripples that distort significantly the free surface and are out of phase with the standing waves they generate; ripples are the triangular shaped bed-forms that have ``lengths of less than about 2 feet and heights of less than about 0.2 foot''. In some other articles (\cite{HSU69,Y77}), the classification of bedforms is rather based on the ratio of the grain diameter $d$ to the viscous sub-layer depth, i.e. on the particle Reynolds number $u_* d/\nu$ ($u_*$ is the shear velocity and $\nu$ the kinematic viscosity of the fluid). These criterions are questionable, especially when there is no clear separation of length-scale between the spacing of ripple crests and the flow depth, and also in the case of bedforms superimposed on larger structures (\cite{VCB05a}). In this paper, willing to refer to the dynamical mechanisms involved in the formation of these bedforms, we shall call ripples those whose characteristics are independent of the flow depth; in contrast, dunes are directly related to the presence of a free surface. In particular, we shall not use the terms `sand wavelets', as suggested by Coleman et al. (\cite{CM94,CM96,CE00,CFG03}) to designate the initial stage of ripples when they emerge from a flat sand bed.

Experimental measurements on subaqueous ripples are numerous and exhibit large data dispersion. One of the reasons for such a dispersion is that ripples exhibit pattern coarsening, i.e. present a progressive increase of their typical length-scale as time goes by (\cite{M78,GS89,B94,CM94,CM96,B99,RU01,CFG03,VCB05b,VL08,RLF08}). Many authors have only measured fully-developed wavelengths, whose relation with the initial wavelength $\lambda$ is still an open issue that will be discussed here. A reference data base of final wavelengths in inclined flumes has been completed by \cite{Y85}. Although not directly comparable to the predictions of a linear stability analysis, these data show distinct scaling laws for hydraulically smooth and rough granular beds. For a particle Reynolds number $u_* d/\nu$ smaller than few unities, the viscous sub-layer is larger than the grain size $d$ and the fully-developed ripple wavelength turns out to scale on $\nu/u_*$, with a large --~yet unexplained~-- prefactor ($\sim 10^3$). On the other hand, when the grain diameter $d$ is larger than the viscous sub-layer, the wavelength was found to slowly increase with $u_*$. Here we will focus on this hydraulically rough regime, in which the flow is turbulent at all scales down to the grain size. This corresponds to the domain of validity of the calculations performed in part I. In several experimental articles (\cite{B94,CM94,CM96,CE00,CFG03,VL05}), the initial wavelength $\lambda$ has been carefully determined, showing scaling laws independent of the flow depth $H$. At large particle Reynolds number, $\lambda$ is found to be almost independent of the shear velocity $u_*$ and to increase with the grain diameter $d$ (for instance \cite{CFG03} propose $\lambda \propto d^{0.75}$). These measurements will serve as a benchmark for the theory developed in the present paper.

Concerning dunes, most of the measurements have been performed in natural rivers, for which the Froude number $\Fr$ is low (see \cite{B05} for a recent review). In the Mississipi river (\cite{H98}), in the Missouri river (\cite{A72}), in the Rhine (\cite{CGOR00,WB03}) and in the Rio Paran\'a (\cite{PBOHKL05}), the observed wavelengths range from $0.5H$ to $20H$ for Froude numbers around $0.2$. The suggestion by \cite{K63}, \cite{A85} or \cite{C04} that mature dunes should present a well selected wavelength scaling on $H/\Fr^{2}$ is thus far from reflecting the natural dispersion of field data. Flume experiments (see \cite{GSR66}, \cite{RU01} and the data collected by \cite{K63}) have been performed in a much larger range of Froude numbers (from $0.1$ to $1$) and also show well-developped bedform wavelengths between $H$ and $30H$. The extensive data published by \cite{GSR66} are particularly impressive, and the interested reader should refer to the original article rather to the truncated data-set plotted by \cite{F74}. We will discuss these data in detail in the last section of the present article.

\begin{figure}
\includegraphics{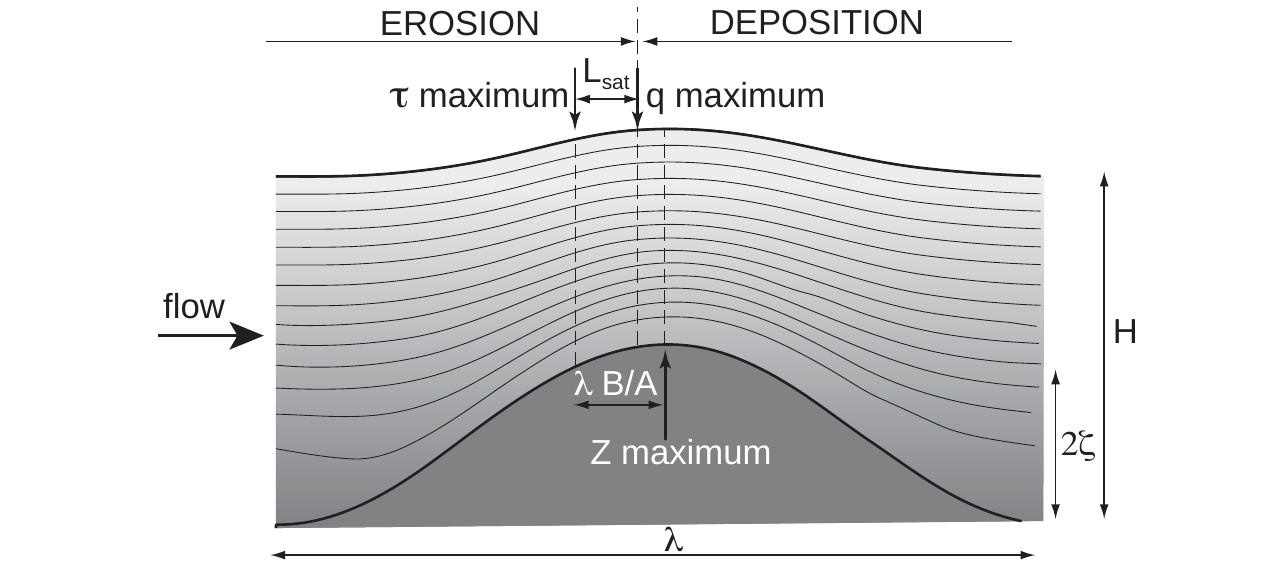}
\caption{Schematic of the instability mechanism showing the streamlines around a bump. The fluid is flowing from left to right. A bump grows when the point at which the sand flux is maximum is shifted upwind the crest. The upwind shift of the maximum shear stress with respect to the crest scales on the size of the bump. The spatial lag between the shear and flux maxima is the saturation length $L_{\rm sat}$.
\label{Schematic}}
\end{figure}

It is widely accepted that ripples result from a linear instability, whose key mechanism is the phase difference between sediment transport and bed topography (\cite{K63,R65,K69,S70,H70,P75,EF82,McL90}). This spatial shift can be conceptually splitted into two contributions. As seen in Part 1, there is a lag between the bed elevation profile and the basal shear stress, encoded in the ratio $B/A$, which results from the simultaneous effects of inertia and dissipation i.e. from hydrodynamics only (Fig.~\ref{Schematic}). In addition, sediment transport needs some time/length to adapt to some imposed shearing. \cite{K63} and his followers (\cite{E70,H70,S70,F74}) have introduced a phenomenological spatial lag $L_{\rm sat}$ (this lag is called $\delta$ in these papers) in a phase shift equation:
\begin{equation}
q(x)=q_{\rm sat} (x-L_{\rm sat}),
\label{shiftedflux}
\end{equation}
where $q_{\rm sat}$ is the saturated (or equilibrium) flux, which is a function of the basal shear stress. With such relation between $q$ and $q_{\rm sat}$, the phase shift between the sediment flux and the shear stress over a relief of wavelength $\lambda$ is then simply $-2\pi L_{\rm sat}/\lambda$, which corresponds to a phase delay for $\lambda>2 L_{\rm sat}$, but to a phase advance for $L_{\rm sat}<\lambda<2 L_{\rm sat}$. As shown by \cite{P75}, such a constant phase lag is not physically founded and should be replaced by a first order relaxation equation of the form
\begin{equation}
L_{\rm sat} \dr_x q = q_{\rm sat} - q.
\label{relaxationflux}
\end{equation}
This linear equation reflects the fact that the sand flux reaches its saturated value over a characteric length $L_{\rm sat}$ (\cite{P75,SKH01,ACD02a,ACD02b,KSH02,VL05,C06}). As for any other first order system, the phase delay between the flux and the shear stress increases here from $0$ for large wavelengths to $\pi/2$ for wavelengths much smaller than $L_{\rm sat}$. When the sum of the phase lags coming from hydrodynamics and transport results into a maximum flux upstream the bed crest, sediment deposition occurs on the bump, leading to an unstable situation, i.e. to the amplification of the bedform (Fig.~\ref{Schematic}). Conversely, when the maximum flux is located downstream the crest, the bump is eroded and the disturbance decay. With the flux equation (\ref{shiftedflux}), even a potential description of the flow (\cite{K63,R65}) for which the basal shear stress is in phase with the topography (see part 1), can then find an unstable -- but unphysical -- region. However, sediment transport has a definite stabilising role, and the destabilising mechanism must result from hydrodynamics.

The description of a turbulent flow over a sinusoidal bottom has been, in this context, progressively improved by \cite{E70}, \cite{F74}, \cite{R80} and \cite{SB84}. None of these articles take into account a saturation length. As discussed in Part~1, the basal shear stress is generically found to be in phase advance with respect to the bottom profile. This means that, without any stabilising mechanism, all wavelengths are found unstable. A key mechanism introduced by \cite{F74} and kept in the following models is the stabilisation of short wavelengths by slope effects. In these models, the prefered wavelength results from hydrodynamics and is proportional to the hydrodynamical roughness $z_0$ (or to $\nu/u_*$ in the case of a viscous surface layer treated by \cite{SB84}). As shown by \cite{C06}, the predicted wavelengths are smaller than all experimental findings by several orders of magnitude.

The concept of flux saturation was correctly introduced in a linear stability analysis by \cite{P75}, the flow being modelled by depth averaged equations standardly used in hydraulics. However, these equations cannot describe the layered structure of the turbulent flow above bumps (see \cite{JH75} and Part 1), and in particular the so-called inner layer which is responsible for the phase shift between shear and topography. As a consequence, this author missed the explaination of the ripple instability. The idea that $L_{\rm sat}$ is the relevant lengthscale for the problem of dune formation was introduced in the aeolian context by \cite{SKH01,ACD02b,KSH02}. Linear stability analyis developed in this context (\cite{ECA05,CA06,AC07}) suggest that the initial wavelength actually scales on $L_{\rm sat}$, and not on $z_0$. This line of thinking has been applied since then to subaqueous ripples (\cite{V05,VL05,C06}). Although $L_{\rm sat}$ and $z_0$ are both ultimately related to the grain size $d$, they correspond to very different mechanisms associated to sand transport and hydrodynamics respectively.

Although \cite{R80} suggested that dunes could form by the same linear instability as ripples, no article has ever exhibited a proper and complete dispersion relation showing, in the same graph, the growth rate $\sigma$ for a range of wave-numbers $k$ that includes both ripples and dunes. In the figures provided by \cite{R80}, $\sigma$ is rescaled by $k^2$, which artificially enhances the small $k$ (dunes) region. More recently, Colombini (\cite{C04,CS05}) revisited this problem, introducing the thickness of the transport layer has a key parameter. This empirical refinement gives in practice an adjustable phase shift between relief and sand transport and allows to get a single well-defined peak in the dispersion relation, associated to dunes, at a wavelength on the order of the flow depth. However, this model does not predict the existence of a small-scale (ripple) instability anymore.

In this paper, we wish to discuss afresh this ripple and dune formation problem, and in particular to get such a complete relation dispersion, where the peak associated to ripples is present. To reach this goal, we must first discuss the sediment transport issue, before mixing it with the hydrodynamics presented in Part~1. In contrast to \cite{R80} and \cite{C04}, we eventually show that dunes cannot form by a linear instability mechanism, and must rather result from the interaction and non-linear pattern coarsening of ripples, limited by the free surface, as suggested by \cite{RW90} and \cite{R06}.

This paper is structured as follows. In the next section devoted to the sediment transport issue, we show how to abstract the details of the transport mechanisms into a general framework for the description of the different modes of transport and thus valid in the subaqueous and aeolian cases. A quantitative model of the transport threshold that includes the slope effects is proposed. We then show that the sediment transport can be limited either by the erosion rate or by the momentum available in the flow. With these ingredients, we propose a self-consistent model that unifies these two mechanisms and discuss the scaling law of the saturation length $L_{\rm sat}$. In section~3, we revisit the formation of current ripples (and aeolian dunes) by linear instability, using the quantitative hydrodynamic calculation performed in part~1. It confirms that the most unstable wavelength is determined by the saturation length, and not by the hydrodynamical roughness. Inverting the problem, this allows to determine the saturation length from experimental measurements of the initial wavelength and to discuss its dependence on the grain size and the shear velocity. Finally, using the systematic expansion of the flow field with respect to the corrugation amplitude performed in part~1, we discuss the non-linear selection of ripple aspect ratio. In section~4, we investigate the effect of a free surface on the linear instability. We show that the excitation of standing waves at the surface has a stabilising effect, independently of the details of the flow and sediment transport models, which is by no means consistent with the formation of river dunes by a primary instability. In the last section, we present field experiments performed in the Leyre river, in France, that directly evidence the formation of ripples by a linear instability and the formation of dunes by a non-linear pattern coarsening limited by the free surface. Further observations suggest that mega-dunes form when the sand bed presents hetereogeneities such as a wide distribution of grain sizes. Finally, we review and discuss the data available in the literature in the light of these theoretical and experimental results.

%________________________________________________________________________
\section{The saturation length paradigm}

%__________________________________
\subsection{Saturation transient}
We present here a general framework to describe, with only few key variables, the sediment transport properties. This framework can accomodate various situations associated to different modes of transport (bed load, saltation, reptation, etc) and different dynamical mechanisms controlling this transport (hydrodynamical erosion of the bed, splash, mixing by turbulent fluctuations, etc).

The reference situation is a uniform turbulent boundary layer of constant shear velocity $u_*$ over a flat sand bed characterised by a threshold shear velocity $u_{\rm th}$. In this situation, one observes a steady uniform transport characterised by a flux $q_{\rm sat}$ called the saturated flux, which corresponds to an equilibrium state between flow and transport. $q_{\rm sat}$ is a function of $u_*$ and $u_{\rm th}$. It is important to emphasise the precise meaning of $u_*$ in this context. On the one hand, $u_*$ is defined from the Reynolds shear stress, i.e. from the averaged correlation of velocity fluctuations $\overline{u'w'}$. On the other hand, the transport is mostly determined by the velocity field inside the transport layer of thickness $h_0$ (few grain diameters in the case of bed-load and typically $1$~cm in the aeolian case). If the latter is embedded in the inner layer, then the average velocity is locally logarithmic and can be related to the basal shear stress. $u_*$ should thus be determined from velocity measurements performed inside this inner layer --~i.e. typically at a distance of $\ell \simeq 10^{-2}\,\lambda$ from the bed, see part 1. The time-scale over which the momentum is exchanged increases linearly with the distance $z$ from the surface (it scales as $z/u_*$). Thus, the velocity fluctuations at time-scales smaller than $z/u_*$ should be associated to the turbulent shear stress and those slower than $z/u_*$, to variations of $u_*(t)$. The shear velocity $u_*(t)$ relevant for sediment transport should thus be averaged over the time-scale $h_0/u_*$.

Considering an inhomogeneous situation --~for instance a spatial variation of the shear stress~-- the sand flux is not instantaneously in equilibrium with the local shear stress. In most of the situations, the transient toward equilibrium can be described by a first order relaxation law, with a single time and length scales:
\begin{equation}
T_{\rm sat} \partial_t q+L_{\rm sat} \dr_x q = q_{\rm sat} - q,
\label{charge}
\end{equation}
where the flow goes in the increasing $x$-direction. In a situation homogeneous in space, $T_{\rm sat}$ is the time needed for the flux to reach again the saturation if the flow speed suddenly changes. Conversely, in a steady situation where there is no sediment flux at the entrance of a volume of control, $L_{\rm sat}$ is the length needed for the flux to reach $q_{\rm sat}$.

As the relaxation time is usually much smaller than the time-scale of evolution of the bedform, we are left with a description of the transport by three variables: the saturated flux $q_{\rm sat}$, the threshold shear velocity $u_{\rm th}$ and the saturation length $L_{\rm sat}$. Note in particular that gravity effects are included into the transport threshold. This multi-scale approach allows to separate clearly the dynamical mechanisms that govern the emergence of bedforms from those governing sediment transport. In other words, subaqueous ripples and aeolian dunes are of the very same nature: a primary instability that is not sensitive at all to the mode of transport (e.g. bed-load \emph{vs} saltation, see \cite{HDA02,CA06}). This framework lies on two important assumptions. First, the depth of the inner layer in which the shear stress is vertically homogeneous should always remain larger than the depth of the transport layer --~see part 1. Second, there is in general more than one relaxation mode and consequently a whole spectrum of relaxation times and lengths. The description by a first order relaxation implies that one of these modes is significantly slower than (and thus dominant in front of) the others. Otherwise, two or more saturation lengths have to be taken into account in a higher order relaxation equation.

An important such case is the suspended load, which must be described by a field (the sediment concentration) and not by a scalar (the integrated flux). Of course, a saturated flux $q_{\rm sat}$ exists in that case (the only particularity is the explicit dependence of $q_{\rm sat}$ on the flow depth) but, there is an infinite number of modes describing the behavior around saturation. We will thus restrict the discussion to the case of grains sufficiently heavy to prevent suspension. This is realized in practice for ripples and dunes in rivers, at moderate Froude numbers. It has been observed by \cite{GSR66} that antidunes, which form in above $\Fr=1$, produce so much turbulent kinetic energy that suspended load could be dominant. This situation, probably highly non-linear, and which requires a specific treatment (see \cite{P75,C04,CS05})), is beyond the scope of the present article.

We aim here to discuss the origin of the existence of a transport saturated state and to propose a self-consistent model, predicting the main characteristics of transport: the static threshold, the saturation length and the saturated flux.

%__________________________________
\subsection{Static threshold}
The bed load threshold is directly related to the fact that the grains are trapped in the potential wells created between neighbouring grains at the sand bed surface. To obtain the scaling law of the threshold shear stress, the simplest geometry to consider is a single spherical grain jammed between the two fixed grains below it (see figure~\ref{Trajectory} in the Appendix~\ref{Staticthreshold}). We consider the situation in which the cohesive forces between the grains are negligible and the friction at the contacts is sufficient to prevent sliding. The loss of equilibrium occurs for a value of the driving force $F$ proportional to the submerged weight of the grain: $F \propto ( \rho_s - \rho_f )g d^3$, where $\rho_s$ is the mass density of the sand grain, and $\rho_f$ is that of the fluid. As $F$ is proportional to the shear force $\tau d^2$ exerted by the fluid, the non-dimensional parameter controlling the onset of motion is the Shields number $\Theta$, which characterises the ratio of the driving shear stress to the normal stress:
\begin{equation}
\Theta = \frac{\rho_f u_*^2}{(\rho_s-\rho_f)gd}.
\end{equation}
The threshold value can be estimated from a force balance. In the viscous regime, the grain drag coefficient $C_d$ is inversely proportional to the grain Reynolds number $\mathcal{R}= U d/\nu$: $C_d=s^2 \mathcal{R}^{-1}$; it is constant ($C_d=C_\infty$) in the turbulent regime. Note that the turbulent drag is not only due to the fluctuations induced by the grain itself, as in the case of a grain falling in a fluid at rest, but also to those present in the turbulent flow. This situation has never been studied properly so far. Note also that the drag force is {\it a priori} different in the vicinity of the ground and that the lift force could be non-negligible when the grain is trapped at the surface of the sand bed. In order to avoid introducing two many parameters, we will not consider these effects here.

\begin{figure}
\includegraphics{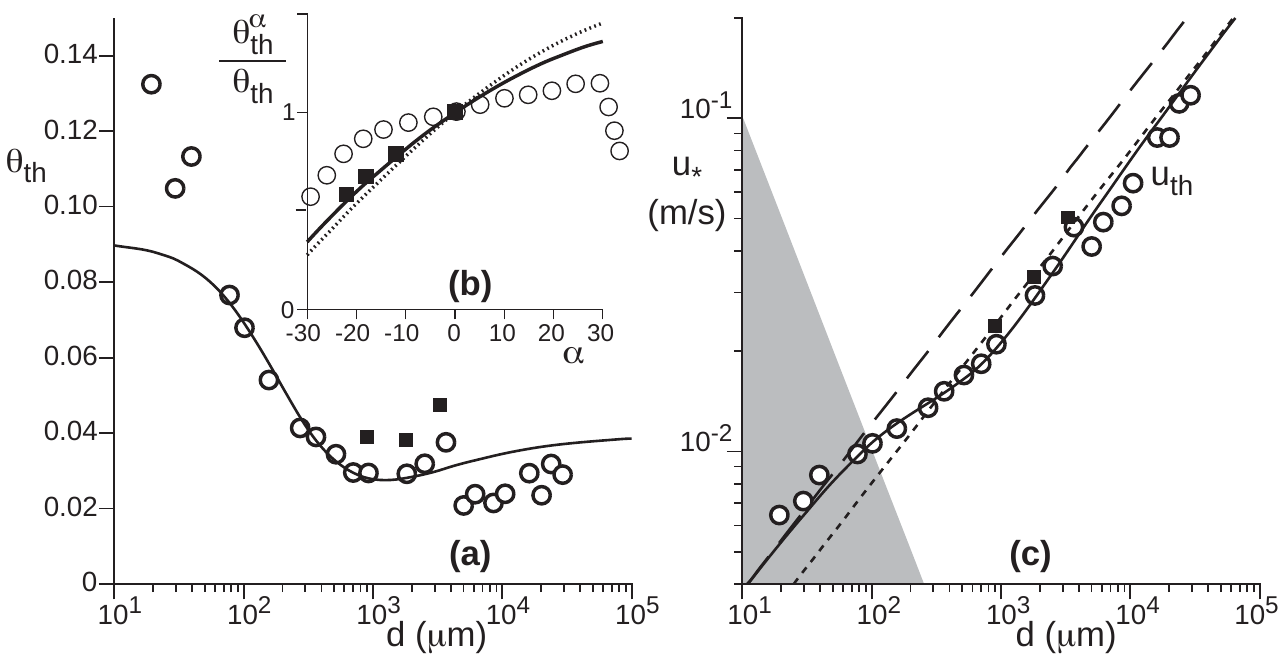}
\caption{(a) Threshold Shields number $\Theta$ as a function of the grain size $d$, for natural sand grains in water. Symbols: measurement by \cite{YK79} ($\circ$) and by \cite{LvB76} ($\blacksquare$). Solid line: Comparison with the model (equations~\ref{uth1}~and~\ref{uth2}). (b) Dependence of the threshold Shields number $\Theta_{\rm th}^\alpha$ on the bed inclination $\alpha$. Symbols: measurements by \cite{LvB76} for natural sand grains ($\blacksquare$) and by \cite{LGRD05} using glass beads ($\circ$) in a narrow channel. Solid line: expectation of the model (equations~\ref{uth1}~and~\ref{uth2}), for natural sand grains. Dotted line: approximation by $\cos \alpha+\sin \alpha/\mu$, with $\mu=\tan 32^\circ$. (c) Threshold shear velocity $u_{\rm th}$ as a function of the grain size $d$, for natural sand grains. Symbols: measurement by \cite{YK79} ($\circ$) and by \cite{LvB76} ($\blacksquare$). Solid line: Comparison with the model. Dotted line: asymptotic behaviour in the turbulent regime. Dashed line: asymptotic behaviour in the viscous regime. Grey zone: zone in which the viscous sub-layer is larger than the bed roughness: for grains smaller than $d=100~\mu$m, the bed is hydraulically smooth for any shear stress $u_*$ above the threshold $u_{\rm th}$.
\label{Threshold}}
\end{figure}

Typical values found in sedimentation experiments performed with natural sand grains are $C_\infty \simeq 1$ and $s \simeq 5$. The generic situation is a uniform shear stress $\rho_f u_*^2$, for which the velocity profile is linear in the viscous regime:
\begin{equation}
u_x=\frac{u_*^2 z}{\nu}
\end{equation}
and logarithmic in the turbulent regime:
\begin{equation}
u_x=\frac{u_*}{\kappa} \ln\left(1+\frac{z}{rd}\right),
\end{equation}
where $\kappa$ is the von K\'arm\'an constant and $r$ the ratio of the aero or hydrodynamic roughness to the grain diameter. The threshold Shields number is constant (see Appendix \ref{Staticthreshold}), both in the viscous
\begin{equation}
\Theta_{\rm th} =\frac{8 \mu}{3 s^2}
\end{equation}
and in the turbulent regimes:
\begin{equation}
\Theta_{\rm th} = \frac{4 \mu \kappa^2}{3 C_\infty \ln^2(1+1/2r)}\, ,
\end{equation}
where $\mu$ is the avalanche slope. The laminar-turbulent transition, indicated by the grey zone in figure~\ref{Threshold}c, is determined by the Reynolds number and thus depends on the grain diameter $d$. In most of the situations relevant in geophysics, $d$ lies in the cross-over between these two regimes. Figure~\ref{Threshold} shows the comparison between experimental data for natural sand grains and the full model, whose derivation is provided in appendix~\ref{Staticthreshold}. There is no adjustable parameter in the model as one can independently measure the avalanche slope $\mu=\tan 32^\circ$, the drag coefficients and the sand bed roughness $r=0.1$.

It is worth noting that continuum mechanics cannot predict correctly this threshold. Assuming that the granular presssure $P$ and the shear stress $\tau$ are continuous at the fluid/grain interface, the standard Coulomb criterion $\tau/P=\mu$ leads to a null threshold shear stress. Part of the problem comes from the fact that the pressure $P$ felt by the first layer of grains is due to its own weight: $\simeq (\rho_s-\rho_f) gd$. Introducing such a pressure discontinuity, one finds a threshold Shields number  $\Theta_{\rm th} \simeq \mu$, which is independent of the flow regime and much too large, compared to measurements. This means that there is no continuity between the fluid and the granular shear stresses, making continuum descriptions of sediment transport inherently problematic.

%__________________________________
\subsection{Gravity or slope effect}
Due to gravity, it is much easier to transport sediments along the lee side of a sand bump than on the stoss slope. This effect is directly encoded into the dependence of the saturated flux on the transport threshold, which depends on the slope. Let us note $\alpha$ the bed angle at the scale of the saturation length. The force balance now reads:
\begin{equation}
\frac{\pi}{6} ( \rho_s - \rho_f ) g d^3 \left(\sin \alpha+\mu \cos \alpha\right) = \frac{\pi}{8} C_d  \rho_f \left(U_{\rm th}^\alpha\,d\right)^2.
\end{equation}
and $U_{\rm th}^\alpha$ is the fluid velocity around the grain at the threshold  (see Appendix~\ref{Staticthreshold}). We keep the notations $u_{\rm th}$ and $\Theta_{\rm th}$ for the threshold shear stress and the threshold Shields number for an horizontal bed ($\alpha=0$) and denote by $u_{\rm th}^\alpha$ and $\Theta_{\rm th}^\alpha$ the thresholds in the case of an inclined bed. 

The dependence of the threshold on the slope has been measured experimentally by \cite{LvB76} for sand grains, in the turbulent regime. The model matches quantitatively the results, without any adjustable parameter (solid line on figure~\ref{Threshold}b). Neglecting the fact that $C_d$ is a function of the grain Reynolds number, the threshold shear stress can be written as: $\rho_f \left(u_{\rm th}^\alpha\right)^2=\rho_f u_{\rm th}^2 (\cos \alpha+\sin\alpha/\mu)$. Inserting this expression into the saturated flux formula, one can see that gravity leads to a down-slope component of the flux, i.e. to a diffusion mechanism. This approximation still gives a reasonable fit of the data (dotted line in figure~\ref{Threshold}b).

The slope dependence of the threshold shear stress has also been measured in the case of aeolian transport, for rough sand grains (\cite{RIR96}) and is again perfectly fitted by the model. However, there exists an unexplained discrepancy in the case of spherical glass beads in a viscous liquid (Fig.~\ref{Threshold}b), as one would need a very low effective friction $\mu \sim \tan 70^\circ$ much lower than the avalanche slope ($\mu \sim \tan 24^\circ$) to fit the data (\cite{LGRD05}). With such a value, the gravity effect becomes completely negligible.

Importantly, the whole gravity effect on sand transport can be encoded into the threshold slope dependence, at the linear order in $\alpha$. This was directly proved experimentally by \cite{LvB76} in the sub-aqueous case and by \cite{RIR96} in the aeolian case. In different articles (\cite{R80} and followers), $\mu$ was considered as a parameter that can be adjusted to fit the initial ripple wavelength. As they  were missing the correct stabilising mechanism, aberrantly low values of the repose angle were found. In section~\ref{wavelengthselection} (see also Fig.~\ref{LambdaMax}), where the wavelength selection issue is discussed, we shall keep for $\mu$ the avalanche slope determined experimentally.

%__________________________________
\subsection{Erosion limited transport}
When a sand bed is submitted to a flow, only a small fraction of the grains at the surface are entrained and this erosion process takes some time to occur. We first model the erosion rate, and in particular the role played by the disorder of the surface. We define $\mathcal{N}(\Theta)$ the fraction of grains at the surface susceptible to be entrained at a Shields number $\Theta$. In the absence of flow, for $\Theta=0$, $\mathcal{N}$ is null. It increases very quickly around the mean threshold Shields number $\Theta_{\rm th}$ and reaches $1$ for large velocities.  $\mathcal{N}(\Theta)$ reflects and encodes the distribution of potential wells at the sand bed surface. Whatever the functional form chosen for this quantity, there are two important parameters: the Shields number around which $\mathcal{N}$ switches from $0$ to $1$, which reflects the overall transport threshold, and the typical slope $\mathcal{N}'$ in the same zone. A simple choice is:
\begin{eqnarray}
\mathcal{N}&=&0\quad {\rm if} \quad \Theta<\Theta_m, \nonumber\\
\mathcal{N}&=&\frac{\Theta-\Theta_m}{\Theta_M-\Theta_m}\quad {\rm if} \quad \Theta_m<\Theta<\Theta_M, \\
\mathcal{N}&=&1\quad {\rm if} \quad \Theta>\Theta_M.\nonumber
\end{eqnarray}
For the sake of simplicity, we consider that one grain at the surface occupies an area $d^2$. We start from a very simple assumption: when one grain is entrained, we hypothesise that there can be no further erosion in the surrounding area until the grain has moved by a distance comparable to its own diameter $d$. The description of the elementary jump by one grain diameter is derived in appendix \ref{grainMotion}. The erosion time ${\mathcal T}$ is the time needed to move over a distance $d$. It behaves as:
\begin{equation}
{\mathcal T} \propto\,\ln \left[\frac{\sqrt{\Theta}+\sqrt{\Theta_{\rm th}}}{\sqrt{\Theta}-\sqrt{\Theta_{\rm th}}} \right]\,\sqrt{\frac{\rho_f d}{(\rho_s-\rho_f) g}}
\label{ApproxTturb}
\end{equation}
in the turbulent regime and
\begin{equation}
{\mathcal T} \propto\,\ln \left[\frac{\Theta+\Theta_{\rm th}}{\Theta-\Theta_{\rm th}} \right]\, \left(\frac{d_\nu }{d}\right)^{3/2}  \,\sqrt{\frac{\rho_f d}{(\rho_s-\rho_f) g}}
\label{ApproxTvisc}
\end{equation}
in the viscous regime. This means that the mean velocity goes to $0$ at the threshold like $1/\ln \left[\Theta-\Theta_{\rm th} \right]$, which corresponds to a bifurcation much sharper than the standard saddle node bifurcation in $\sqrt{\Theta-\Theta_{\rm th}}$. The experimental observation (\cite{AF77,LvB76,CME04}) that the grain velocity --~i.e. the mean velocity conditioned by the fact that the grain moves significantly at the time resolution of the instrument~-- is non zero just above the threshold is thus fully consistent with this expression. Far from threshold, the grain velocity is just proportional to the flow velocity (\cite{AF77,LvB76,CME04}).
 
The erosion flux $\varphi_\uparrow$ is the volume of grains eroded per unit time and unit surface. It can be expressed as an integral over all possible local configurations:
\begin{equation}
\varphi_\uparrow= \frac{\pi d^3}{6 \phi} \int_0^{\Theta}  \frac{\mathcal{N}'\left(\Theta_{\rm th}\right)}{d^2\,T(\Theta_{\rm th},\Theta)} d \Theta_{\rm th},
\end{equation}
where $\phi$ is the bed volume fraction. A good approximation of this integral is given by:
\begin{equation}
\varphi_\uparrow \propto \frac{\left( \Theta-\Theta_m \right)}{\phi \, \left(\Theta_M-\Theta_m \right)\,\left[c_m+\ln \left(\frac{\Theta+\Theta_m}{\Theta - \Theta_m}\right)\right]} \,\sqrt{\left(\frac{\rho_s}{\rho_f}-1\right) g d}\quad {\rm for} \quad \Theta_m<\Theta<\Theta_M,
\end{equation}
where $c_m\simeq1.41$ in the turbulent regime and $c_m\simeq0.605$ in the viscous regime. In particular one can recover from this expression the expected asymptotic behaviours.
Close to the threshold, there is a regime in which the density of mobile grains is not sufficient to induce a significant reduction of the flow strength in the transport layer, and the grains can be considered as isolated, which corresponds to a dilute transport layer. The saturation of the flux is then controlled by the erosion rate and the disorder of the sand bed. This idea is due to \cite{CME04}, who proposed a semi-phenomenological model that uses extensively experimental results. In the present model, the length of a trajectory is independent of the trap in which the grain was initially at rest and fluctuates around a well defined average value ${\mathcal L}$. The mean area explored by the grain is ${\mathcal L} d$ and contains, by definition of ${\mathcal L}$, a mean number of potential wells sufficiently deep to trap the grain equal to $1$. We then obtain the expression:
\begin{equation}
{\mathcal L}=\frac{d}{1-\mathcal{N}(\Theta)} \, .
\label{LsatAqueux}
\end{equation}
At small $\Theta$, ${\mathcal L}$ is simply one grain diameter $d$; at large $\Theta$, $\mathcal{N}$ tends to $1$ so that ${\mathcal L}$ diverges. This divergence has been directly evidenced experimentally in the viscous case, by \cite{CLDZ08}. In that case, the measured value of $\Theta_M$ is around $2\Theta_m$. ${\mathcal L}$ is the length needed for the transport to adapt to a change of shear stress. It thus gives the saturation length $L_{\rm sat}$.

This behaviour constrains the domain  of validity of this regime: the shear velocity should be around the threshold, the transport should be intermittent and diluted and the distribution of potential wells should be very large. As suggested by the experiments of \cite{CME04}, a granular bed prepared by sedimentation is initially very disordered and consequently presents a wide range of potential wells. It then takes a very long time for the surface to re-arrange, leading to a drift of the distribution $\mathcal{N}$ toward larger and larger threshold shear stresses. This long transient is probably not relevant to geophysical situations in which the time and length scales of the systems are always sufficiently large to ensure that an equilibrium state for the geometrical arrangement of surface grains has been reached. The saturated flux is simply the erosion rate times the hop length ${\mathcal L}$:
\begin{equation}
q_{\rm sat}\propto \frac{\left( \Theta-\Theta_m \right)}{\phi \, \left(\Theta_M-\Theta \right)\,\,\left[c_m+\ln \left(\frac{\Theta+\Theta_m}{\Theta - \Theta_m}\right)\right]} \,\sqrt{\left(\frac{\rho_s}{\rho_f}-1\right) g d^3}\quad {\rm for} \quad \Theta_m<\Theta<\Theta_M.
\label{eqqqqmoinsun}
\end{equation}
By construction, it vanishes at the Shields number $\Theta_m$ and diverges at $\Theta_M$. The transport is thus limited by the erosion rate only close to the threshold $\Theta_m$, in the presence of a disordered bed. As the density of transported grains becomes important, the transport becomes limited by the available momentum.

\begin{figure}
\includegraphics{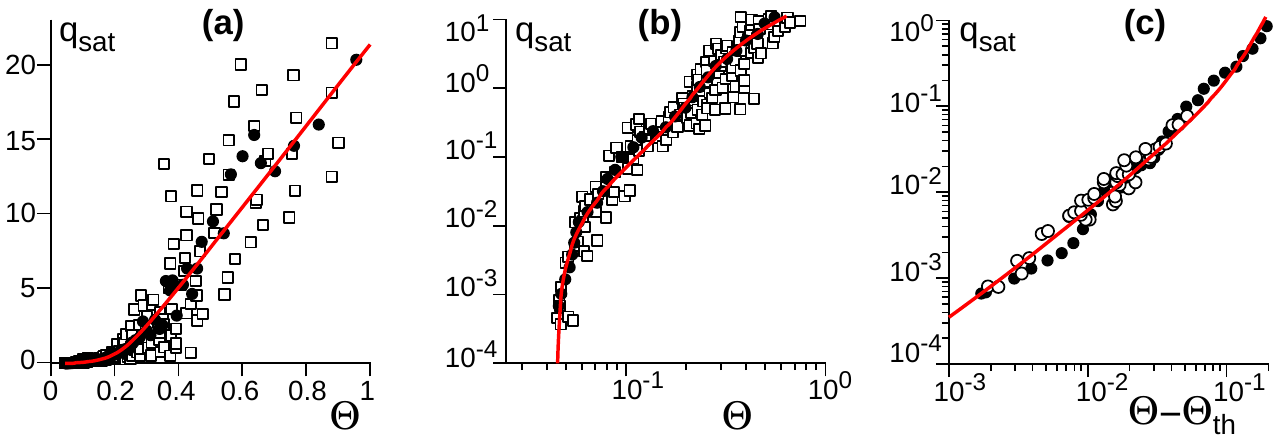}
\caption{Saturated flux $q_{\rm sat}$ rescaled by $\sqrt{\left(\frac{\rho_s}{\rho_f}-1\right) g d^3}$ as a function of the Shields number $\Theta$: raw data ($\square$) gathered by \cite{J98}; same after local averaging ($\bullet$); raw data ($\circ$) obtained by \cite{LvB76}. The solid line is the best fit by the model described here (equations \ref{qqq1} and \ref{qqq3}), which gives: $\Theta_{\rm th}=0.045$ and $\Theta_{\rm M}=5\,\Theta_{\rm th}$. (a) Lin-lin plot; (b) Log-log plot; (c) Log-log plot as a function of  $\Theta-\Theta_{\rm th}$.
\label{FluxSature}}
\end{figure}

%__________________________________
\subsection{From erosion limited to momentum limited transport}
One can extend the previous model by including the negative feedback of the transport on the flow (\cite{B56}). Each time the flow entrains a grain from the bed and accelerates it, the grain exerts in return a stress on the fluid. The balance between erosion and deposition is the same as in the previous paragraph (Eq.~\ref{eqqqqmoinsun}) except that the shear stress in the transport layer is reduced to some value $u_{\rm f}$, compared to that in a particle-free flow:
\begin{equation}
q_{\rm sat}\propto \frac{\left( u_{\rm f}^2-u_{\rm th}^2 \right)}{\phi \, \left(u_{\rm M}^2-u_{\rm f}^2 \right)\,\left[c_m+\ln \left(\frac{u_{\rm f}^2}{u_{\rm f}^2-u_{\rm th}^2} \right)\right] }\,\sqrt{\left(\frac{\rho_s}{\rho_f}-1\right) g d^3}.
\label{qqq2}
\end{equation}
We consider the regime in which the grains are transported at the surface of the bed i.e. do not form a surface sheet flow. They leave the bed with a velocity $v_\uparrow$ and collide back the sand bed with a velocity $v_\downarrow$. The sand-borne shear stress is proportional to the sand flux and to the difference ($v_\downarrow-v_\uparrow$). The fluid in the transport layer is assumed to be in equilibrium between the driving shear stress $\rho_f u_*^2$, the fluid-borne basal shear stress $\rho_f u_{\rm f}^2$ and the sand-borne shear stress:
\begin{equation}
\rho_f u_*^2=\rho_f u_{\rm f}^2+ \rho_s \phi \frac{(v_\downarrow-v_\uparrow)}{{\mathcal L}} \, q.
\label{qq0}
\end{equation}
This relation is nonetheless valid in the saturated state ($q=q_{\rm sat}$) but also during the transient of saturation. It is very important to understand that the particles are entrained by the flow at a velocity reduced by the presence of other particles. The grain trajectory is encoded into a single quantity, $(v_\downarrow-v_\uparrow)/{\mathcal L}$, which is a function of the reduced shear velocity $u_{\rm f}$ and \emph{not} of $u_*$. For the sake of simplicity, we limit the discussion to the turbulent case. The transposition to the viscous case is straightforward. Assuming that $v_\downarrow-v_\uparrow$ scales on the grain mean velocity, the ratio $(v_\downarrow-v_\uparrow)/{\mathcal L}$ is proportional to the hop time ${\mathcal T}$, which is also the erosion time. Equation~(\ref{qq0}) can be solved to get the flux:
\begin{equation}
q_{\rm sat} \propto \frac{\rho_f}{\rho_s}\, \left(u_*^2-u_{\rm f}^2\right)\,T(u_{\rm f}) \propto \frac{\rho_f}{\rho_s}\,\sqrt{\frac{\rho_f d}{(\rho_s-\rho_f) g}} \, \ln \left[\frac{u_{\rm f}+u_{\rm th}}{u_{\rm f}-u_{\rm th}} \right] \, \left(u_*^2-u_{\rm f}^2\right).
\label{qqq1}
\end{equation}

Eliminating the flux between the equations (\ref{qqq1}) and (\ref{qqq2}), one obtains a relation between $u_{\rm f}$ and $u_*$.
\begin{equation}
\left(u_*^2-u_{\rm th}^2\right) \propto \left(u_{\rm f}^2-u_{\rm th}^2 \right)\left[1+ \frac{\frac{\rho_s}{\rho_f}\,\left(\frac{\rho_s}{\rho_f}-1\right) g d}{\phi \, \left(u_{\rm M}^2-u_{\rm f}^2 \right)\,\left[c_m+\ln \left(\frac{u_{\rm f}^2}{u_{\rm f}^2-u_{\rm th}^2} \right)\right]\,\ln \left[\frac{u_{\rm f}+u_{\rm th}}{u_{\rm f}-u_{\rm th}} \right] }\right].
\label{qqq3}
\end{equation}
Just above the threshold, the flow in the transport layer is undisturbed: $u_f \sim u_*$. The erosion limited regime is recovered and the flux can be approximated by:
\begin{equation}
q_{\rm sat}\propto \frac{\left( u_*^2-u_{\rm th}^2 \right)}{\phi \,\left[c_m+\ln \left(\frac{u_*^2}{u_*^2-u_{\rm th}^2} \right)\right]}\,\sqrt{\left(\frac{\rho_s}{\rho_f}-1\right) g d^3}
\end{equation}

Far above the threshold, for large values of $u_*$, the shear velocity $u_f$ inside the transport layer  tends to $u_{\rm M}$ i.e. to a value independent of $u_*$.  The flux then scales as:
\begin{equation}
q_{\rm sat} \propto \frac{\rho_f}{\rho_s}\,\sqrt{\frac{\rho_f d}{(\rho_s-\rho_f) g}} \, \left(u_*^2-u_{\rm M}^2\right)
\end{equation}
Thus, in the momentum limited regime, $q$ scales at large velocities as $u_*^2$ and not $u_*^3$ as usually obtained. The same result is valid for aeolian transport, for which the negative feedback of particles on the flow has been directly evidenced experimentally. Although many authors have followed Bagnolds, the scaling law in $u_*^2$ is, in that case, the best description of existing data (\cite{A04}). In figure~\ref{FluxSature}, we have re-plotted the measurements of the saturated flux collected by \cite{J98} for the subaqueous case. It exhibits a large dispersion due to several factors. First, data coming from systems with different grain size distributions have been plotted together without any distinction of symbols. Second, the reproducibility of experiments is made difficult by the problem of granular bed preparation: we emphasise again that sedimentation leads to an out of equilibrium situation that can last for days. Third, it is difficult to estimate the basal shear stress in flume experiments due to lateral boundaries. We have also plotted the measurements performed by \cite{LvB76}, which are much less scatterred. Given these reservations, the fit by the model derived here provides a good description of existing data. In particular, the asymptotic behaviors close to the threshold and for large shear velocities are well captured.

As many other formulas would provide an equally good fit (i.e. \cite{MM48}, \cite{E50}, \cite{B56} or \cite{Y63}), forthcoming sediment tranport dedicated experiments should be designed to test directly the ingredients of the models. For instance, the measurement of the hydrodynamic roughness in presence of transport can provide a direct evidence of the negative feedback of particles on the flow (see \cite{A04} for the aeolian case). One expects this roughness to be larger than in the transport-free case as a consequence of the increase of the dissipation rate -- the fact that the particles are moving does not lead to a drag reduction.

The last key issue is the saturation length. The sand flux is the product of the grain velocity by the density of mobilised grains. Two important mechanisms can thus control the saturation length: the length needed by the grain to reach its asymptotic velocity --~the so-called drag length~-- and the length needed for erosion to take place i.e. the trajectory length ${\mathcal L}$. The modelling of the drag length is a difficult problem as the trajectory takes place in a turbulent flow whose fluctuations are \emph{not} due to the motion of the grain itself. The problem is thus very different from that of a sphere moving in a fluid at rest, a problem for which the drag law is calibrated. To the best of our knowledge, the motion of a sphere whose diameter lies in the inertial range of the turbulent flow is still an open problem. We are thus left with the standard drag force formula $\frac{\pi}{8} C_d \rho_f U^2 d^2$, with a drag coefficient of order one. Then, solving the equation of motion, one obtains a drag length $L_{\rm drag}$ around $2\,(\rho_s/\rho_f)d$. In summary, we consider here that $L_{\rm sat}$ can be either limited by erosion, in which case it is expected to scale as:
\begin{equation}
L_{\rm sat} \simeq \frac{\Theta_M-\Theta_m}{\Theta_M-\Theta}\,d
\end{equation}
or by grain inertia, in which case it should scale as:
\begin{equation}
L_{\rm sat} \simeq 2 \frac{\rho_s}{\rho_f} d
\end{equation}

One can see that these two predictions are difficult to test and discriminate. The erosion length gently increases with the shear velocity while the drag length is independent of it; the drag length increases with the density ratio $\rho_s/\rho_f$ but, experimentally, this parameter cannot be easily varied by a large factor. We shall see below that the study of subaqueous ripples can shed some light on this issue.

\begin{figure}
\includegraphics{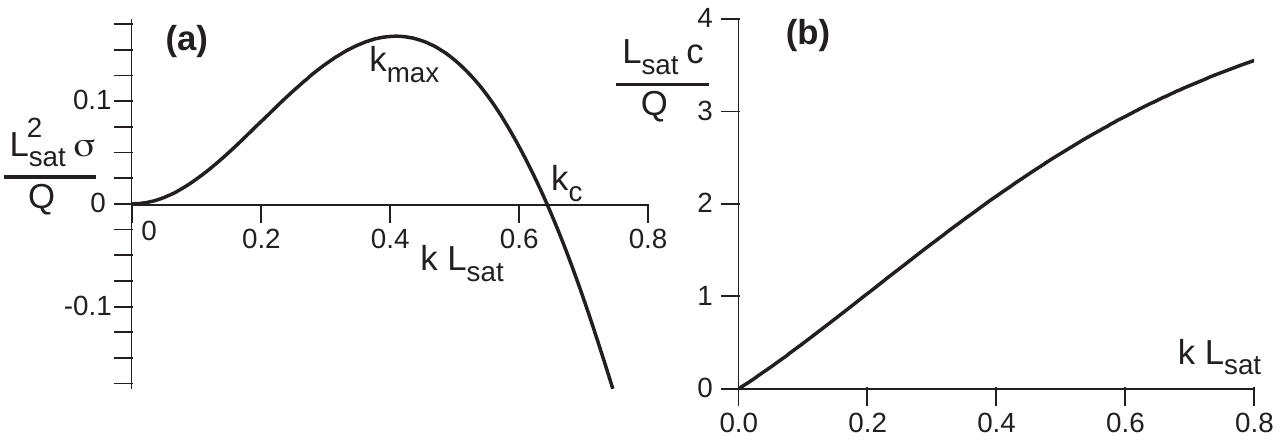}
\caption{Dispersion relation obtained in the erosion limited regime, for $\mu=\tan 32^\circ$ and $u_{\rm th} \ll u_*$. (a) Growth rate $\sigma$ as a function of the wave-number rescaled by the saturation length: $kL_{\rm sat}$. (b) Propagation speed $c$. $Q$ is the reference flux (Eq.~\ref{Qerosionlimited}).}
\label{DispersionRelation}
\end{figure}
%

%________________________________________________________________________
\section{Subaqueous ripples}

%__________________________________
\subsection{Linear stability analysis}
Given that the details of the sand transport only affect the saturated flux, the saturation length and the threshold shear stress, we can perform the linear stability analysis of a flat sand bed in a very general framework. We consider a periodic disturbance of the bed profile $Z$. As the base state is homogeneous, we can seek for modes of the form $exp(\sigma t + i k (x-ct))$. The shear stress $\sigma_{xz}$ induced by the wavy sand bed is written in the Fourier space as
\begin{equation}
{\hat \sigma}_{xz} = u_*^2 (A + iB) k \hat{Z},
\label{S_t}
\end{equation}
where $u_*$ is the shear velocity in the reference state and $k=2\pi/\lambda$ the wave number associated to the spatial coordinate $x$. We refer the reader to the part 1 of this article for the derivation of the components of the shear stress $A$ and $B$ respectively in phase and out of phase with respect to the bottom profile. The lag between the maximum shear stress and the ripple crest is given by $\lambda B/ (2\pi A)$. The computation of the coefficients $A$ and $B$ has been performed under the assumption that the sand bed can be considered as static. As a matter of fact, the growth rate $\sigma$ is related to the sediment transport and on the order of $q_{\rm sat}/L_{\rm sat}^2$. Experimentally, $\sigma$ is usually found to be four orders of magnitude smaller than the typical flow shear rate: the time-scale of formation of subaqueous ripples is typically ten seconds to be compared to few tens of microseconds for the period at which the flow is excited by the dune relief, $\lambda/U$. Therefore, in contrast to what is suggested by \cite{CS05}, the normal velocity of the grain-fluid interface does significantly change the flow in this problem.

The threshold shear stress is a function of the slope:
\begin{equation}
\hat u_{\rm th}^\alpha=u_{\rm th} \frac{ i k \hat{Z}}{2 \mu},
\end{equation}
where $\mu$ is the avalanche slope. At linear order, we can write the saturated flux $q_{\rm sat}(u_*,u_{\rm th})$ as:
\begin{equation}
\hat q_{\rm sat}=\frac{\partial q_{\rm sat}}{\partial u_*} \frac{u_*}{2} (A + iB) k \hat{Z}+\frac{\partial q_{\rm sat}}{\partial u_{\rm th}} \frac{u_{\rm th}}{2 \mu} i k \hat{Z}
\end{equation}
Introducing a reference flux $Q$, we rewrite this expression under the form:
\begin{equation}
\hat q_{\rm sat}=Q (a+ib) k \hat{Z}
\end{equation}
where $a$ and $b$ are the components of the saturated flux in and out of phase with the topography. Now the flux relaxes to its saturated value with a spatial lag $L_{\rm sat}$:
\begin{equation}
ik L_{\rm sat} \hat q=\hat q_{\rm sat}-\hat q
\end{equation}
Using the conservation of matter $\partial_t Z + \partial_x q=0$, one obtains the dispersion relation:
\begin{equation}
\sigma-ikc=-ik \frac{\hat q}{\hat Z}=-\frac{ik}{1+ik L_{\rm sat} }\frac{\hat q_{\rm sat}}{\hat Z}=-\frac{i Q (a+ib) k^2}{1+ik L_{\rm sat} }
\end{equation}
Splitting the equation into its real and imaginary parts, one obtains the growth rate $\sigma$ and the propagation speed $c$:
\begin{eqnarray}
\frac{L_{\rm sat}^2 \sigma}{Q}&=&\frac{(kL_{\rm sat})^2 (b - a kL_{\rm sat})}{1+(kL_{\rm sat})^2}\label{sigma} \\
\frac{L_{\rm sat} c}{Q}&=&\frac{(kL_{\rm sat}) (a + b kL_{\rm sat})}{1+(kL_{\rm sat})^2} \label{cc}
\end{eqnarray}
This corresponds to a standard convective instability at large wavelengths (Fig.~\ref{DispersionRelation}). The cut-off wave-number $k_ c$ above which modes are stabilised by the saturation length is given by:
\begin{equation}
k_c L_{\rm sat} =\frac{b}{a} \, .
\end{equation}
Note that the instability can present a different threshold that that for the transport if $b$ vanishes at some value of $u_*$ larger than $u_{\rm th}$. In first approximation, $a$ and $b$ are weak functions of $kz_0$. Then, one can approximate the maximum growth rate wave-number $k_{\rm max}$ as:
\begin{equation}
k_{\rm max} \, L_{\rm sat} \simeq X^{-1/3}-X^{1/3} \;\;\; {\rm with}\;\;\; X=-\frac{b}{a} + \sqrt{1+ \left(\frac{b}{a}\right)^2}.
\label{kmax}
\end{equation}
As the instability is convective, we have also computed the spatial growth rate. Its maximum nicely coincides with that of the time growth rate.

It can be inferred from the conservation of matter that the propagation velocity $c$ is proportional to the difference of flux $\delta q$ between trough and crest and invertly proportional to the ripple height $2 \zeta$ --~this is the so-called Bagnold relation. At large wavelength $\lambda$, $\delta q$ is proportional to the reference flux $Q$ and to the height so that the propagation speed varies as: $c\propto Q/\lambda$. This is confirmed by figure~\ref{DispersionRelation}b, which shows a roughly linear relation between $c$ and $k$.

%__________________________________
\subsection{Erosion limited transport}
We can make the previous general results more precise by making explicit the expressions for the length $L_{\rm sat}$ and time $L_{\rm sat}^2/Q$ scales, and for a articular relation between sediment flux and shear stress. In the case of the erosion limited transport, the saturated flux reads:
\begin{equation}
q_{\rm sat}= \chi \frac{\left( u_*^2-u_{\rm th}^2 \right)}{c_m-\ln \left(1-\frac{u_{\rm th}^2}{u_*^2} \right)} \, .
\end{equation}
Defining the reference flux as
\begin{equation}
Q = \chi \frac{u_*^2}{c_m-\ln \left(1-\frac{u_{\rm th}^2}{u_*^2} \right)} \, ,
\label{Qerosionlimited}
\end{equation}
we get:
\begin{eqnarray}
a&=& A \left [1 + \frac{(u_{\rm th}/u_*)^2}{c_m-\ln \left(1-(u_{\rm th}/u_*)^2\right)}\right ],\\
b&=& B \left [1 + \frac{(u_{\rm th}/u_*)^2}{c_m-\ln \left(1-(u_{\rm th}/u_*)^2\right)}\right ]-\frac{(u_{\rm th}/u_*)^2}{\mu}\left[1 + \frac{1}{c_m-\ln \left(1-(u_{\rm th}/u_*)^2 \right)}\right].\nonumber
\end{eqnarray}
Close to the threshold, (for $u_* \simeq u_{\rm th}$), $a$ tends to $A$ and $b$ to $B-\mu^{-1}$. For asymptotically large shear stresses i.e. $u_* \gg u_{\rm th}$,  $a$ tends to $A$ and $b$ to $B$. Thus,  the ratio $b/a$ increases with the flow strength. This means that the destabilising mechanism becomes more efficient as $u_*$ increases so that the maximum growth rate wavelength $\lambda_{\rm max}$ decreases with $u_*$ (Fig.~\ref{SigmaModelMu}).

If $B$ is below $\mu^{-1}$ then $b$ vanishes for a value of $u_*$ larger than $u_{\rm th}$. The instability threshold is then distinct from the transport threshold, as observed in the viscous regime. As expected for such an instability, the most unstable wavelength then diverges at the instability threshold (Fig.~\ref{SigmaModelMu}b) --~ in practice, it would be limited by the geometrical size of the experiment. In the turbulent regime, however, $B$ is sufficiently large to ensure that flat beds are unstable as soon as transport takes place. Correspondingly, $\lambda_{\rm max}$ remains finite at $u_*=u_{\rm th}$.
\begin{figure}
\includegraphics{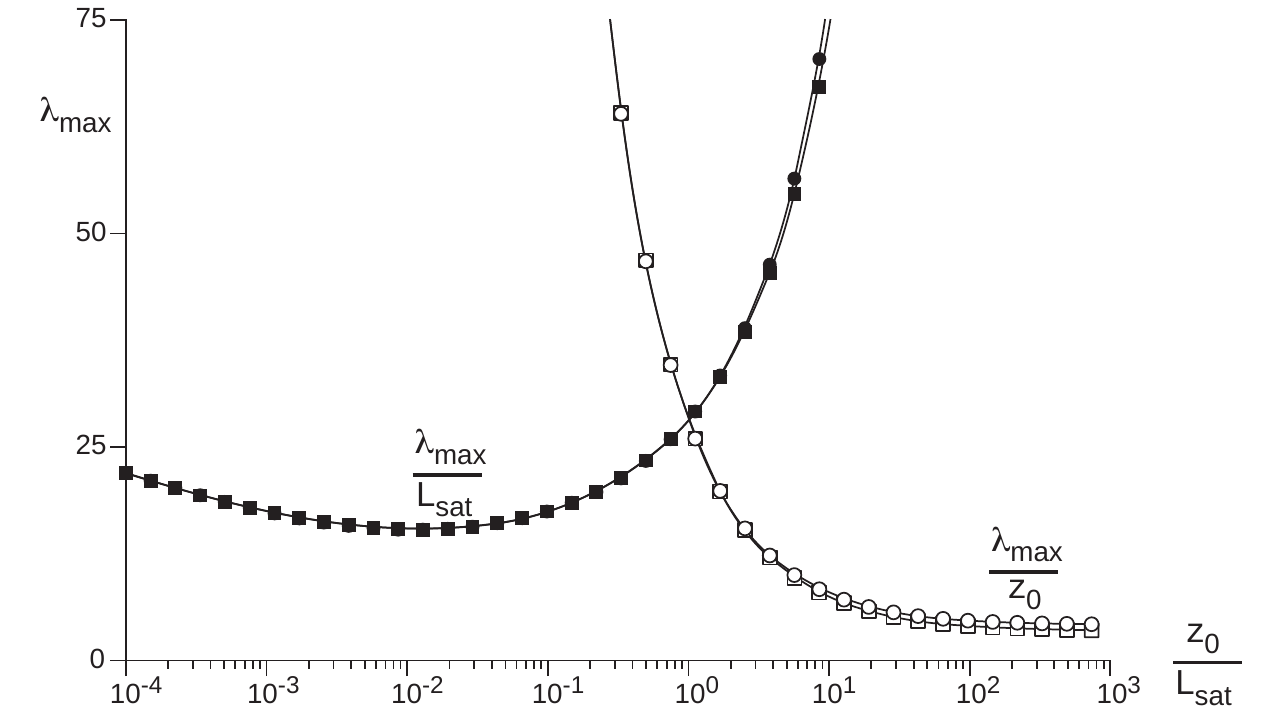}
\caption{Wavelength $\lambda_{\rm max}$ of maximum growth rate as a function of the ratio of the two characteristic lengthscales $z_0/L_{\rm sat}$, in the limit $u_*\gg u_{\rm th}$, for $\mu=\tan 32^\circ$. Black symbols: ratio $\lambda_{\rm max}/L_{\rm sat}$. Open symbols: ratio $\lambda_{\rm max}/z_0$. Squares: erosion limited model. Circles: momentum limited model (One can hardly see differences between the two transport models).
\label{LambdaMax}}
\end{figure}
%

%__________________________________
\subsection{Momentum limited transport}
When the transport is limited by momentum, we can write the saturated flux under the generic form:
\begin{equation}
q_{\rm sat}=\chi u_*^{2\gamma} (u_*^2-u_{\rm th}^2).
\end{equation}
The self-consistent model derived above gives $\gamma=0$, while the \cite{B56} formula involves an exponent $\gamma=1/2$.  Other empirical models such as \cite{MM48}, \cite{E50} or \cite{Y63} can also be approximated in this way. We now define the reference flux $Q$ by:
\begin{equation}
Q= (\gamma+1) \chi u_*^{2 (\gamma + 1)},
\end{equation}
so that $a$ and $b$ read:
\begin{eqnarray}
a&=&A-\frac{\gamma A}{1+\gamma}\,\frac{u_{\rm th}^2}{u_*^2} \, ,\nonumber\\
b&=&B-\frac{\gamma B+\mu^{-1}}{1+\gamma}\,\frac{u_{\rm th}^2}{u_*^2} \, .
\end{eqnarray}
For asymptotically large shear stresses i.e. $u_* \gg u_{\rm th}$,  $a$ tends to $A$ and $b$ to $B$, independently of $\gamma$ and $\mu$.  Close to the threshold, (for $u_* \simeq u_{\rm th}$), $a$ tends to $A/(1+\gamma)$ and $b$ to $(B-\mu^{-1})/(1+\gamma)$. So, again, the instability threshold is above the transport threshold when $B$ is smaller than $\mu^{-1}$.
\begin{figure}
\includegraphics{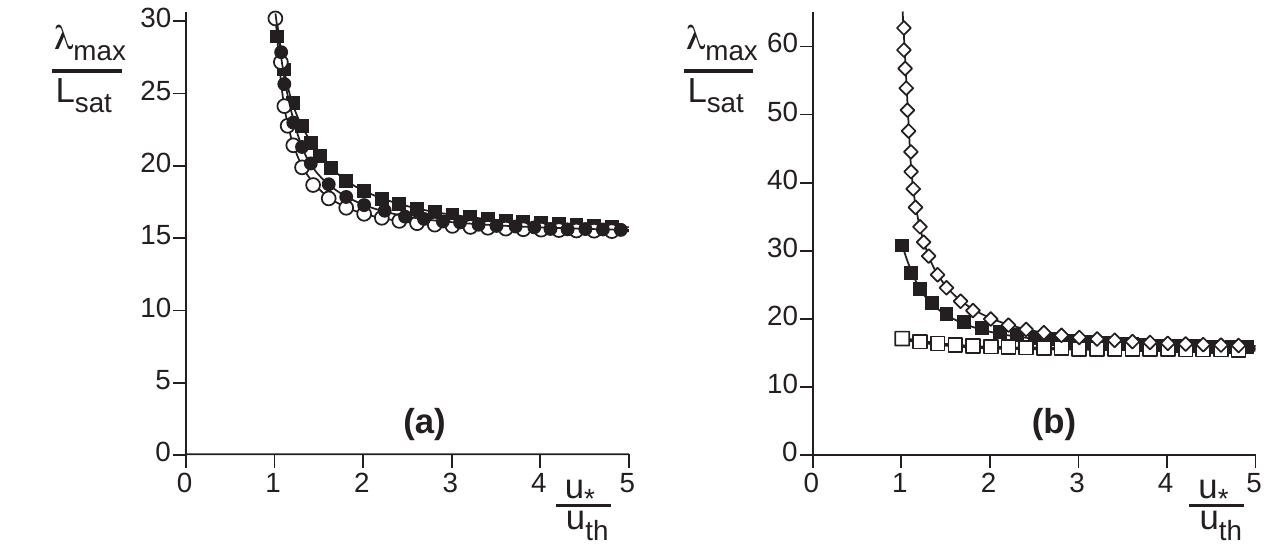}
\caption{Wavelength $\lambda_{\rm max}$ rescaled by $L_{\rm sat}$ as a function of the rescaled shear velocity $u_*/u_{\rm th}$, for $\mu=\tan 32^\circ$ and $L_{\rm sat}/z_0 = 80$ (typical subaqueous case with sand grains). Erosion limited model ($\blacksquare$), momentum limited model ($\gamma=0$, {\large $\circ$}) and \cite{B56} model ($\gamma=1/2$, {\large $\bullet$}). (b) Same plot, computed with the erosion limited model for different values of $\mu$: $\tan 24^\circ$ ({\large $\diamond$}), $\tan 32^\circ$ ($\blacksquare$), $\tan 70^\circ$ ($\square$).
\label{SigmaModelMu}}
\end{figure}
%

%__________________________________
\subsection{Ripples wavelength selection}
\label{wavelengthselection}
Using the results of part 1, the fastest growing wavelength $\lambda_{\rm max}$ can be computed, taking into account the dependencies of $A$ and $B$ on $kz_0$. The most important question is the scaling law followed by this wavelength. There are two length-scales in the problem: $z_0$ which is an hydrodynamical quantity and $L_{\rm sat}$ which is related to the sand transport. Figure~\ref{LambdaMax} shows that the saturation length controls the scaling of $\lambda_{\rm max}$ if $L_{\rm sat}$ is larger than $z_0$, whatever the transport model. Conversely, $z_0$ controls the scaling of $\lambda_{\rm max}$ if it is larger than $10~L_{\rm sat}$. This could be the case in the hydraulically smooth regime for which $z_0$ scales on $\nu/u_*$ (\cite{SB84}). For subaqueous ripples, the measurements of initial wavelength are usually larger than $100~d$ (see Fig.~\ref{Lambda}), while $z_0$ is on the order of $0.1~d$. If the scaling was controlled by $z_0$, one would underestimate the most unstable wavelength by two orders of magnitude. As a consequence, the wavelength is controlled by the transport saturation length $L_{\rm sat}$. This means that models in which a univoque relation between the actual flux and the shear stress is used cannot capture correctly the physics of ripple instability. The second conclusion is that, although the saturation length may be determined by different dynamical mechanisms, aeolian dunes are of the same nature as subaqueous ripples. It means that different modes of sediment transport (saltation, reptation bed-load, etc) in different situations (viscous, turbulent, etc) can lead to bedforms instabilities of same nature\footnote{Of course, the aeolian ripples do not belong to the same class of bedforms as they result from a screening instability (\cite{B41,A87,An90, ACP06}), not from a hydrodynamical instability.}.

Figure~\ref{SigmaModelMu}a shows that the transport model as a negligible influence on the selected wavelength. This evidences that $L_{\rm sat}$ and $q_{\rm sat}$ are the single two relevant quantities encoding the sediment transport details. Due to the stabilising role of gravity encoded in the slope dependence of the threshold, $\lambda_{\rm max}$ increases close to the threshold shear stress. Of course, this effect is very sensitive to the value of $\mu$, as shown in figure~\ref{SigmaModelMu}b. For natural sand grains ($\mu=\tan 32^\circ$), the wavelength  $\lambda_{\rm max}$ decreases from $30~L_{\rm sat}$ at the threshold to $20~L_{\rm sat}$ far from it. For glass beads, the prediction is very different depending whether one takes the avalanche slope ($\mu=\tan 24^\circ$) or the experimental data of \cite{LGRD05} (equivalent to a fictitious $\mu=\tan 70^\circ$) for the slope effect. In the later case, $\lambda_{\rm max}$ is almost independent of $u_*$ whereas it diverges at the threshold in the former case as $B$ is then of the order of $1/\mu$.
\begin{figure}
\includegraphics{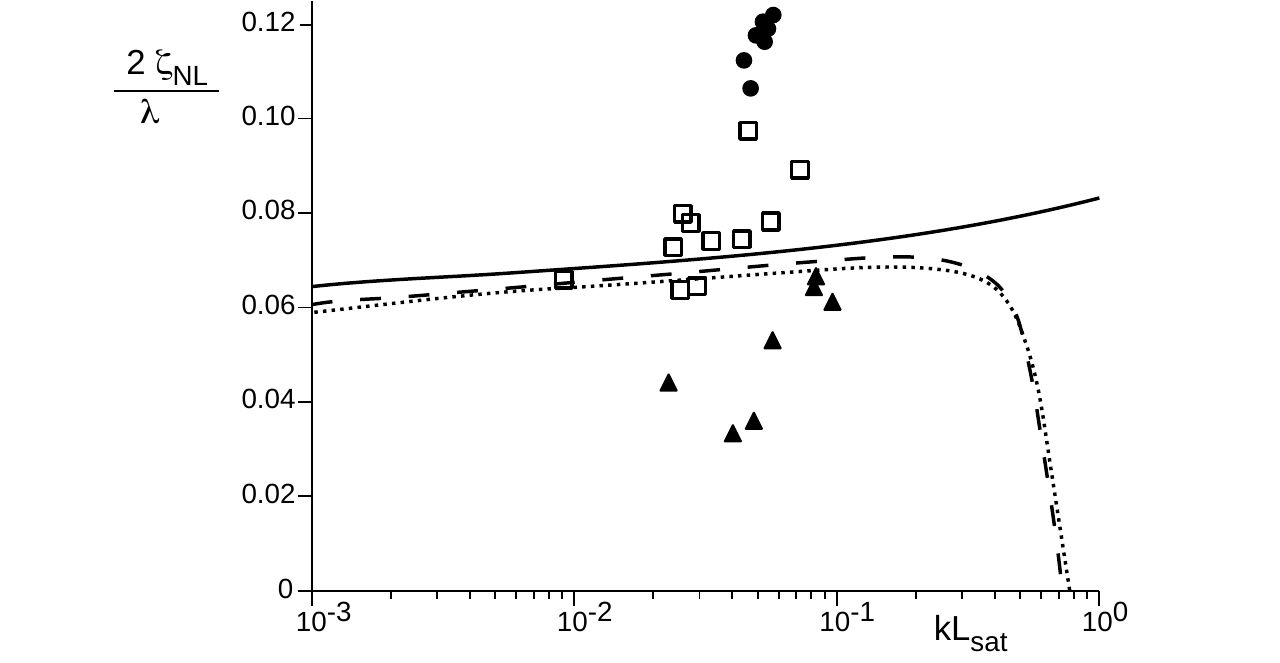}
\caption{Ripples aspect ratio $2 \zeta_{\rm NL}/\lambda$ selected by hydrodynamical non-linear effects as a function of the wave-number $k$ rescaled by $L_{\rm sat}$. The dotted line corresponds to a calculation based on the momentum limited transport model with $L_{\rm sat}/z_0 = 350$ (typical aeolian case), $u_*/u_{\rm th}=3.8$, $\mu=\tan32^\circ$ and $\gamma=0$. The dashed line corresponds to the erosion limited transport model with the same parameters except $L_{\rm sat}/z_0 = 80$ (typical subaqueous case). The solid line does not take any saturation length into account and serves as a reference: the sharp decrease of the two other curves are associated with the stabilisation of small wavelength by the saturation length. The symbols correspond to subaqueous measurements performed by \cite{B99} ($\bullet$) and \cite{GSR66} ($\blacktriangle$) in flume experiments and to aeolian field measurements performed by \cite{PSHMM06,BLW07} ($\square$).}
\label{AspectRatioNonLineaire}
\end{figure}
%

%__________________________________
\subsection{Ripple amplitude selection}
After the linear stage during which the ripples emerge, they exhibit pattern coarsening by progressive merging of bedforms.  During this slow process, each of these ripples may be considered as purely propagative structures that do not grow nor shrink. In the course of their evolution, they should thus present an amplitude $2\zeta$ which is only function of their wavelength $\lambda$.  

In part 1, we have performed a weakly non-linear expansion of the hydrodynamics above a topography up to the third order in $k\zeta$, which is the small parameter of the problem. As a main output of these non-linear calculations, the phase shift between the elevation profile and the basal shear stress decreases as the amplitude $\zeta$ is larger, which stabilises the bedform. Considering a sinusoidal profile, the Taylor expansion of the basal shear stress components takes the form $A=A_1+A_3 (k \zeta)^2$ and  $B=B_1+B_3 (k \zeta)^2$. We denote $\zeta_{\rm NL}$, the particular amplitude for which a purely propagative solution is selected, i.e. which does not grow nor decay: $\partial_t Z=-c \partial_x Z$. This gives the condition: $b = a kL_{\rm sat}$.

Figure~\ref{AspectRatioNonLineaire} shows the prediction of the bedform aspect ratio $2 \zeta_{\rm NL}/\lambda$ as a function of $k L_{\rm sat}$ for two sets of parameters, which corresponds to typical aeolian and subaqueous situations. At small wave-number, the aspect ratio increases logarithmically with $k$. It then decreases abruptly and vanishes at the cut-off wave-number $k_c$. For the aeolian conditions, the predicted dune aspect ratio is around $1/15$ which quantitatively matches field measurements  ($\simeq 1/13$). For current ripples, the measured aspect ratio are much more dispersed: \cite{A85} and \cite{CM94} report values around $1/20$ whereas \cite{B99} found larger values around $1/8$. These discrepancies may be due to the measurement techniques related to the small amplitude of subaqueous bedforms (few tens of grain diameters).
\begin{figure}
\includegraphics{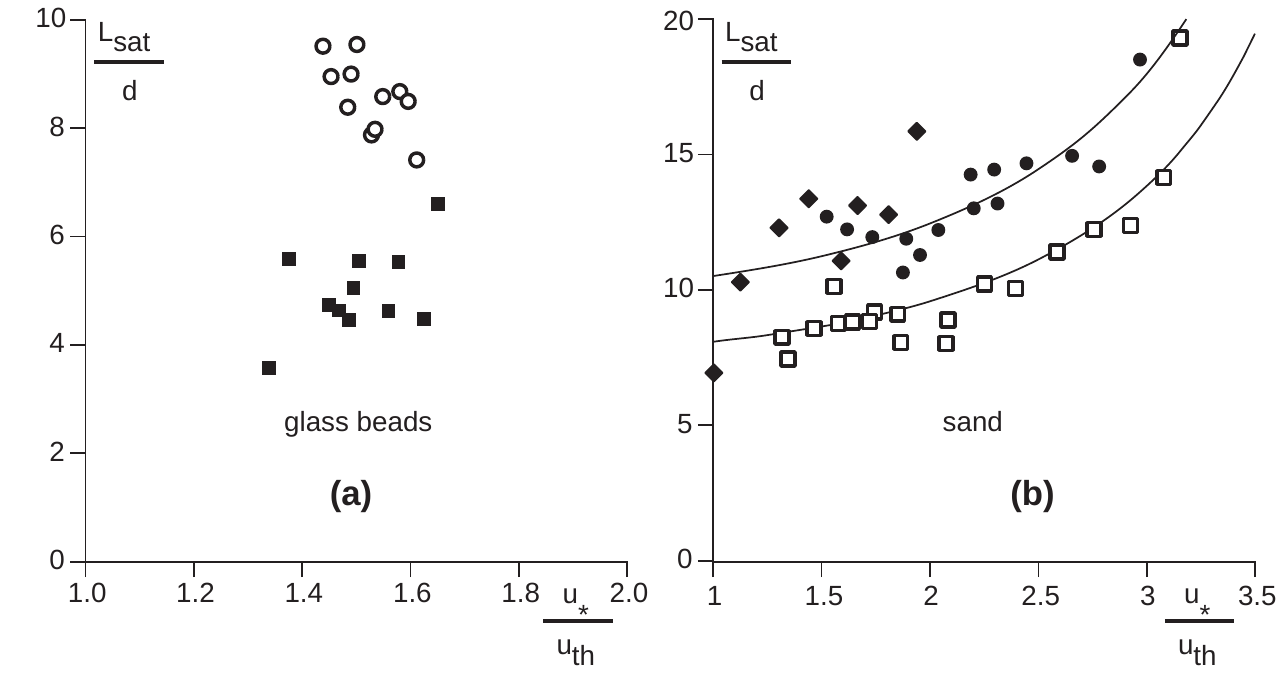}
\caption{Saturation length in water determined from experimental measurements of the initial wavelength as a function of the shear velocity for different types of particles. (a) Glass beads (\cite{VL05}): $d=250~\mu$m ({\large $\bullet$}) and $d=500~\mu$m ($\square$). (b) Natural sand grains (\cite{CM96} and \cite{B99}): $d=210~\mu$m ({\large $\bullet$}), $240~\mu$m ($\blacklozenge$), $d=830~\mu$m ($\square$). The solid lines correspond to the best fit of equation~(\ref{LsatAqueux}). These curves tend to diverge for $u_*=u_M \simeq 4.5~u_{\rm th}$ or equivalently for $\Theta=\Theta_M \simeq 20~\Theta_{\rm th}$. In (a) and (b), the small factor between small and large grains could be due to a subdominant dependence of $L_{\rm sat}$ on viscosity.
\label{LSat}}
\end{figure}

%__________________________________
\subsection{Comparison with experiments}
Rather than predicting the wavelength and comparing it with experimental data, we can invert the process and determine the value of the saturation length that would give the observed wavelength. To play this game, we have chosen the erosion limited transport model and checked that other choices do not change the conclusions reached here. $\mu$ has be chosen equal to the avalanche slope. Most of the experimental data available in the literature correspond to well developed ripples. As these bedforms exhibit pattern coarsening, it is very important to focus on papers reporting the initial wavelength (linear regime), measured for grain sizes larger than $200~\mu$m (hydaulically rough sand bed). We have selected five such data sets: \cite{CM96,B99,VL05}.

As shown in figure~\ref{LSat}, they present consistent trends, which is not the case for the data obtained with smaller grains ($d \sim 100~\mu$m). The saturation length is found to be on the order of several grain diameters. It is slightly smaller for the glass bead experiment than for the natural sand grain ones. The data points around the threshold are very sensitive to the values taken for $\mu$ and $u_{\rm th}$ in the model and should not be over-interpreted. The slight increase of $L_{\rm sat}$ with $u_*$ is more robust, although it is based on the last few data points. More significant is the decrease of the ratio $L_{\rm sat}/d$ when $d$ increases, a feature present for both glass beads and sand grains. It could be related to a subdominant dependence on viscosity.

\begin{figure}
\includegraphics{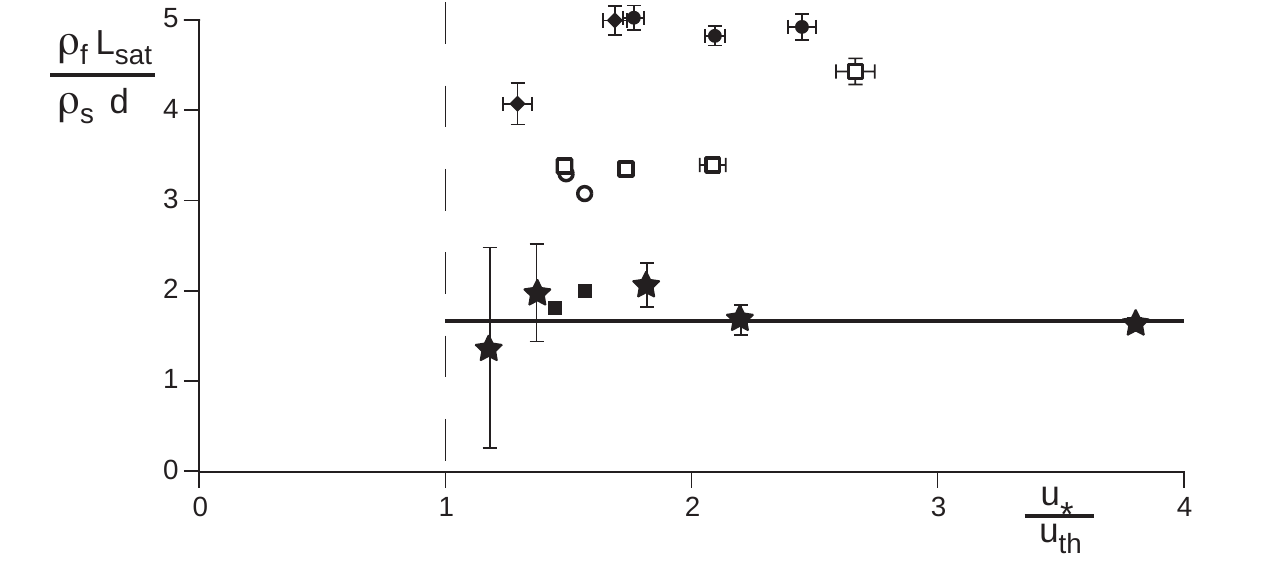}
\caption{Comparison between the saturation length, rescaled by the drag length, in the aeolian ($\bigstar$) and subaqueous (other symbols, see Fig.~\ref{LSat}) cases. For aeolian dunes, $L_{\rm sat}$ is determined from the most unstable wavelength under well characterised winds \cite{ACP08}. For subaqueous ripples, each point of this graph corresponds to the very same data as in figure~\ref{LSat}, but averaged over $6$ measurements. The dashed line corresponds to the entrainment threshold. The solid line is the average over the different points measured in the aeolian case: $L_{\rm sat} \simeq 1.66~(\rho_s/\rho_f)\,d$.
\label{LsatEolien}}
\end{figure}

In the section devoted to the description of sand transport, we have discussed the two simplest possibilities for the dynamical mechanisms limiting the transport saturation: erosion and inertia. Concerning erosion, the prediction $L_{\rm sat}={\mathcal L} \propto d/(1-\mathcal{N}(\Theta))$ is almost impossible to verify as the distribution of potential wells at the surface of the bed is not known. Still, with the simple parametrisation chosen above, one expects the saturation length to scale on the grain size and to gently increase with $u_*$. The solid line in figure~\ref{LSat}b is the best fit by such a form. One obtains the estimate: $\Theta_M/\Theta_{\rm th}\sim 20$. This is far above the value found in figure~\ref{FluxSature}, around $5$, and that found experimentally by \cite{CLDZ08}, around $2$. It means that the erosion length is probably not the mechanism limiting saturation. Otherwise, one would expect a much more rapid increase of the saturation length. Instead, one can observe that it is almost constant, with a subdominant, slow increase with $u_*$.

Figure~\ref{LsatEolien} aims to test the second simple possibility: an inertia limited saturation length. In this case, one expects a scaling law of the form $L_{\rm sat}\propto (\rho_s/\rho_f)\,d$ similar to that observed for aeolian sand transport (\cite{HDA02,A04,CA06,AC07}). We have used the measurements of the wavelength at which aeolian dunes form reported in \cite{ACP08}, obtained either in the field or using aerial photographs. Using the inversion method proposed here, the saturation length  $L_{\rm sat}$ has been obtained for different values of the wind shear velocity (Fig.~\ref{LsatEolien}) and is mostly independent of it. Once rescaled by $\rho_s/\rho_f\,d$, $L_{\rm sat}$ is of the same order of magnitude for both aeolian dunes and subaqueous ripples. In first approximation, all series of data, considered separately, are independent of $u_*$. A better agreement with aeolian data is observed for glass beads and for large grains. A discrepancy by a factor of two is observed for small natural sand grains. As a possible interpretation, the grains roll on the ground during their phase of acceleration, which may lead to underestimate the length needed to reach the fluid velocity. One expects rough sand grains to be more sensitive to this effect.

\begin{figure}
\includegraphics{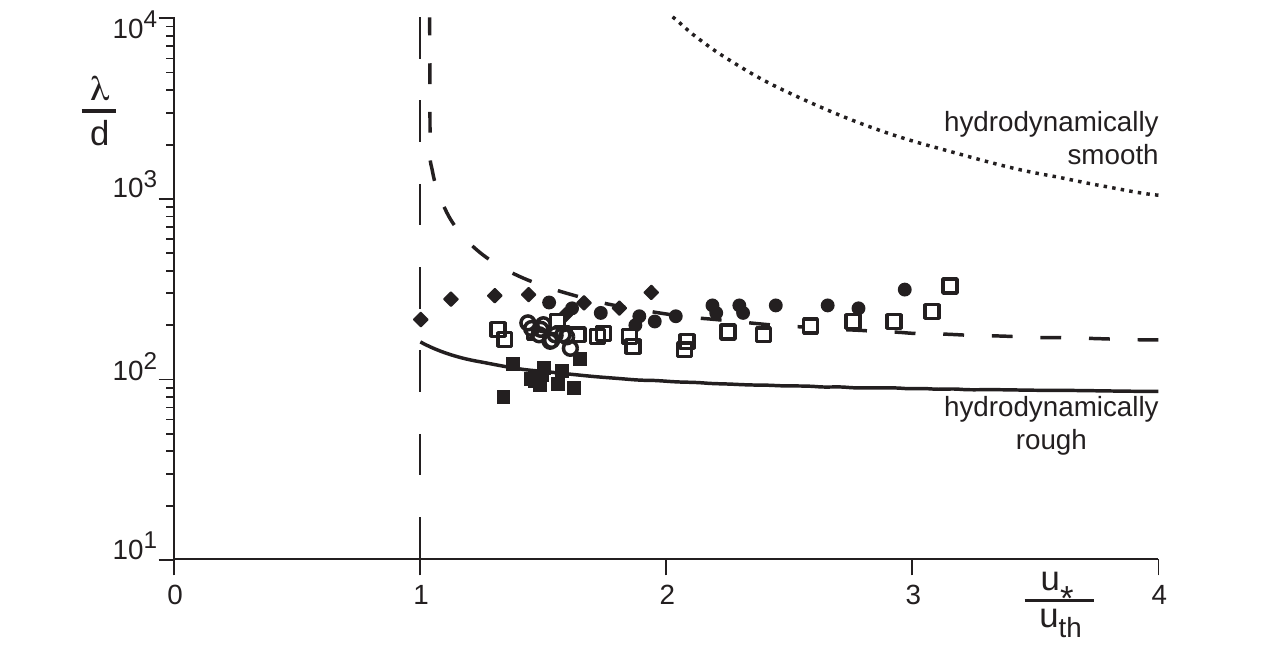}
\caption{Initial wavelength $\lambda$ in water as a function of the shear velocity for different types of particles: glass beads (\cite{VL05}): $d=250~\mu$m ($\bullet$) and $d=500~\mu$m ($\blacksquare$); natural sand grains (\cite{CM96} and \cite{B99}): $d=210~\mu$m ({\large $\bullet$}), $240~\mu$m ($\blacklozenge$), $d=830~\mu$m ($\square$). The lines show the prediction of the model, assuming that the saturation length is given by the drag length, with an hydrodynamically rough sand bed (solid line), with a viscous surface layer (dotted line, $\Rey_t=125$), and in an intermediate regime (dashed line, $\Rey_t=10$).
\label{Lambda}}
\end{figure}

For a given type of grain, the extra-dependence of the initial wavelength on the grain diameter may be attributed to viscous effects, either on the saturation length or on hydrodynamics. To test this, we have plotted in figure~\ref{Lambda} the measurements of the initial wavelength together with different hydrodynamical calculations, assuming that the saturation length is limited by inertia: $L_{\rm sat}$ is kept fixed and equal to $2\,(\rho_s/\rho_f)\,d$. As the shift between the shear stress and the topography gets reduced for a viscous surface layer (see part 1), the predicted wavelength is larger by more than one order of magnitude. Moreover, the coefficient $B$ becomes smaller than $1/\mu$ so that the instability threshold, at which the initial wavelength diverges, is above the entrainment threshold. This applies to grain sizes much smaller than $100~\mu$m (or to more viscous liquids), in the hydrodynamically smooth regime. Still, it should give the trend for the data analysed here: as the grains size decreases, the influence of viscosity increases and so does the initial wavelength. Note that this trend is in the same direction as the empirical observation of \cite{CFG03} that $\lambda/d \propto d^{-0.25}$
%

%________________________________________________________________________
\section{Effect of the free surface}

%__________________________________
\subsection{Dispersion relation}
In the previous section, both hydrodynamic and erosion aspects have been gathered to study bedforms (ripples) under an infinite depth assumption. However, rivers have a free surface and the water depth $H$ is finite. We expect these two additional ingredients to significantly change the shape of the dispersion relation.

\begin{figure}
\includegraphics{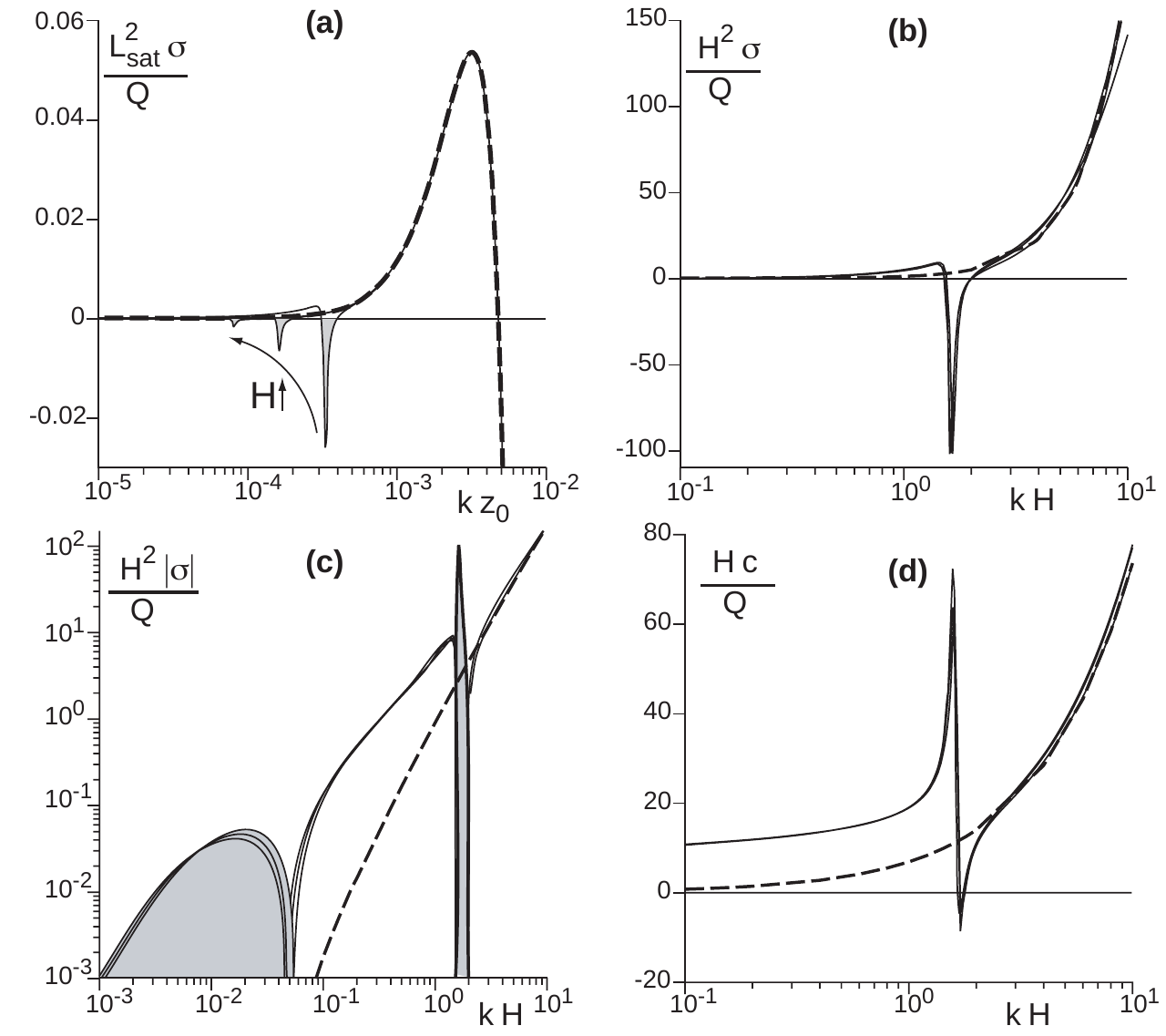}
\caption{Growth rate $\sigma$ and propagation velocity $c$  as functions of the wave-number $k$ for different values of $H/z_0$ ($H/z_0 = 5000$, $10000$, $20000$) and the typical values of parameter $\Fr = 0.8$, $\mu = \tan 32^\circ$, $u_{\rm th}/u_* = 0.8$ and $L_{\rm sat}/z_0 = 80$. In (a), $\sigma$ is rescaled by $L_{\rm sat}$ and $k$ by $z_0$, whereas in (b), (c) and (d) all lengths are rescaled by the relevant length-scale $H$, so that all curves collapse in region where $kH \simeq 1$. The dashed line represents the reference unbounded case presented in the previous section. In panel (c), due to the log-scale, the absolute value of $\sigma$ is displayed and the grey areas encode for negative values of the growth rate.}
\label{LimouSigma}
\end{figure}

\begin{figure}
\includegraphics{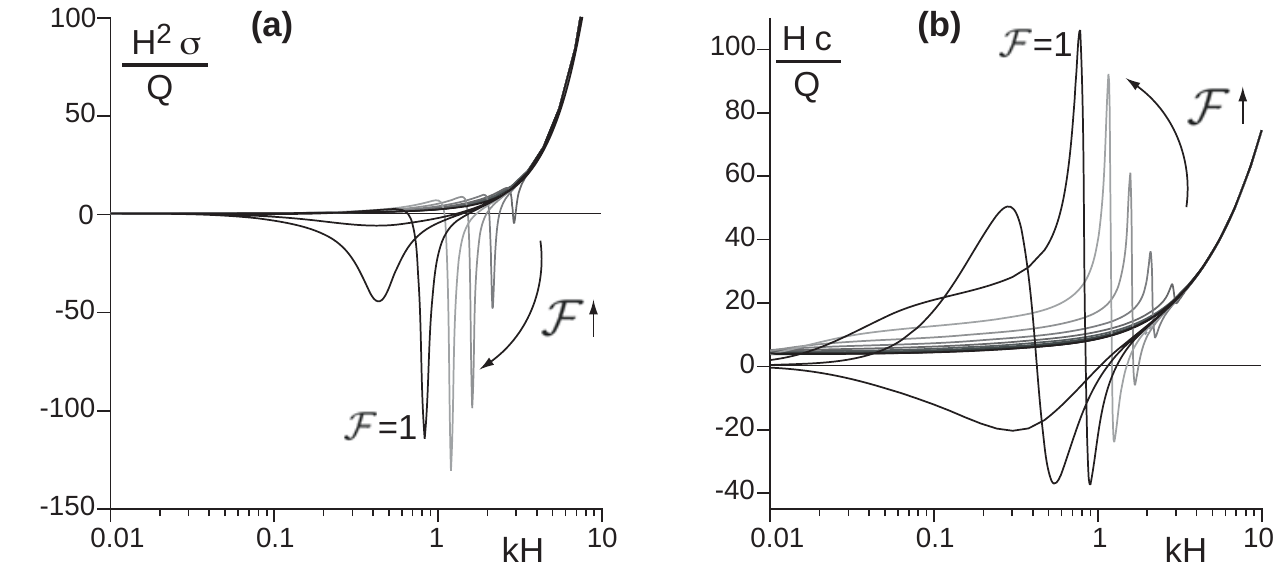}
\caption{Dispersion relation obtained for a ratio $H/z_0 = 10000$, $\mu = \tan 32^\circ$, $u_{\rm th}/u_* = 0.8$ and $L_{\rm sat}/z_0 = 80$. (a) Growth rate $\sigma$ as a function of the wave-number for different values of the Froude number. (b) Propagation speed $c$. In both panels, the Froude number $\Fr$ is varied from $0.5$ to $1.2$ by increment of $0.1$.}
\label{DispersionDunes}
\end{figure}

Our goal is to relate the formation of ripples and dunes to the two relevant length scales $H$ and $L_{\rm sat}$. In part 1, we have precisely computed the basal shear stress coefficients ($A$ and $B$) in the case of a turbulent flow over a wavy bottom. The large wavenumbers are insensitive to the free surface (Fig.~14 in part~1). By contrast, $A$ and $B$ display a resonance peak around $kH \simeq 1/\Fr^2$ and have a divergent behavior as $k \to 0$. As $\sigma$ and $c$ are directly related to $A$ and $B$ through equations~(\ref{sigma}) and (\ref{cc}), they both present the same features. In figure~\ref{LimouSigma}a, the growth rate is represented as a function of $kz_0$ for three different values of $H/z_0$. All curves collapse in the large-$k$ region on the dispersion relation computed in the reference unbounded case. In particular they exhibit a maximum for the same wave-number, which corresponds to the initial ripple wavelength. This means that the presence of the free surface does not influence the formation of ripples as long as $H$ and $L_{\rm sat}$ are well-separated length-scales. In a zone around $kH \simeq 1/\Fr^2$ (see Fig.~\ref{LimouSigma}b), the function $\sigma(k)$ presents, in comparaison to the reference case, a sharp dip which can be attributed to a resonance of gravity surface waves, independently of the transport issue (see part~1). As shown in figure~\ref{DispersionDunes}a, the width and the amplitude of this dip is very sensitive to the value of the Froude number. For small $\Fr$, the effect of the free surface is marginal and the dip is very small. As the Froude number increases, the dip becomes more pronounced so that the growth rate $\sigma$ becomes negative in an enlarging range of wave-numbers: the free surface stabilises wavelengths commensurable with the flow depth. Last, the semi-logarithmic plot of figure~\ref{LimouSigma}c reveals the behavior of the growth rate in the small-$kH$ limit: $\sigma(k)$ tends to $0$ from below. This indicates that the very large wavelengths are also stabilised. In the intermediate range of wavelengths, a slight increase of the growth rate is observed.

The other output of the linear stability analysis is the propagation velocity of the pattern $c(k)$ (Fig.~\ref{LimouSigma}d), which is also very sensitive to the resonance at $kH \simeq 1/\Fr^2$. It presents a sharp maximum on the left of the resonance followed by a dip on the right of it. At this dip, $c$ can become negative for sufficiently large Froude numbers ($\Fr \gtrsim 0.7$).

%

%__________________________________
\subsection{Dune formation}
\begin{figure}
\includegraphics{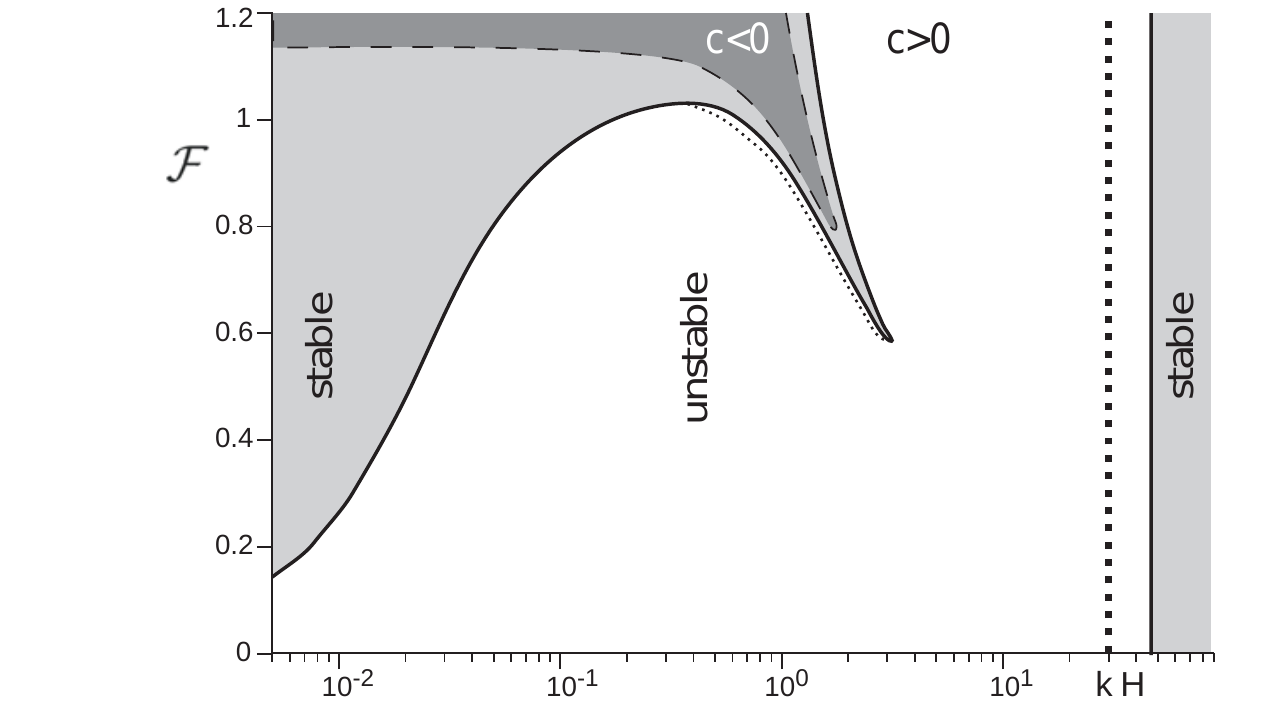}
\caption{Stability diagram parametrised by the Froude number $\Fr$ and the rescaled wave-number $kH$. The marginal stability curves ($\sigma=0$), shown in solid lines, separates the unstable zone (white) from the stable ones (grey). The overall maximum growth rate (bold dotted line) is always reached for ripples. The local maximum of the growth rate resulting from the resonance of standing surface waves is shown in thin dotted line. The dashed line ($c=0$) separates upstream ($c<0$, dark grey) from downstream propagating bedforms. The other parameters are set to: $H/z_0 = 10000$, $\mu = \tan 32^\circ$, $u_{\rm th}/u_* = 0.8$ and $L_{\rm sat}/z_0 = 80$.}
\label{FrkH}
\end{figure}

Gathering the different dispersion relations in the $(\Fr,kH)$ space, one produces the stability diagram (Fig.~\ref{FrkH}). The central region delimited by the two marginal stability curves corresponds to the zone of unstable wavelengths ($\sigma>0$). Large $k$ are stabilised by the saturation length, which explains the lack of dependence on the Froude number in this zone: both the marginal stability and the maximum growth rate curves are vertical lines in the diagram. This simply reflects the fact that ripples do not feel the free surface: they disturb the flow over a thickness of the order of the wavelength $\lambda$ i.e. much smaller than $H$. As already mentioned, the most unstable mode always corresponds to ripples (dotted vertical line in the large-$k$ zone of figure~\ref{FrkH}).

A second zone of stable modes is located at much smaller $kH$ and is associated to the influence of the free surface. The bedforms excite gravity surface waves that propagate at a speed $\sqrt{(g/k)\,\tanh(kH)}$ in the frame of reference moving with the surface particles at the velocity $u_{\rm surface}$. When the upstream wave velocity balances the downstream surface velocity, for 
\begin{equation}
\Fr\sim\sqrt{\frac{\tanh(kH)}{kH}},
\end{equation}
the disturbances induced by the bottom topography accumulate. In these resonant conditions, the wave amplitude reaches a maximum  (Fig.~20b in part~1). The waves are in phase for wave-numbers above the resonance and in antiphase below it (Fig.~20c in part~1). At the resonance, the free surface is in quadrature with the bottom, which tends to move downstream the point of maximum shear stress i.e. to stabilise the bedform. Of course, as the influence of the free surface on the flow is localised over a typical distance $\lambda$, its effect is more important as $kH$ becomes smaller. In summary, there are two conditions under which the free surface can overcome the inertial destabilising effect: (i) around the resonance, since the standing wave amplitude is very large and (ii) in the limit of small wavenumbers, as $H$ becomes much smaller than wavelength $\lambda$. This is precisely what can be observed in figure~\ref{FrkH}, the sharp stable zone surrounding the resonant curve. For obvious reasons, this new minimum of the growth rate associated to the surface wave resonance comes with a local maximum of $\sigma$ (thin dotted line along the restabilised zone). In the work of \cite{R80}, the latter has been associated to the formation of dunes by a linear instability. In many other linear analysis (\cite{K63,R65,E70,F74,CF00,C04,CS05}), the prediction of ripples is missed for the reasons detailed above (the sediment transport relaxation is not correctly taken into account), i.e. the corresponding maximum in the dispersion relation is absent. As a consequence, they are left with a unique peak in the region of $kH$ around unity, which is found here to be a secondary maximum. As explained below, we fundamentally disagree with the conclusion reached in all these papers that subaqueous dunes result from the linear instability of a flat bed.

Let us briefly recall the basic reasons for which one usually associates the appearance of a pattern to a maximum of the growth rate. One considers that the initial condition $Z(t=0)$ is essentially flat, with some wide-band noise. Its Fourier transform $\hat Z(k,t=0)$ then contains some energy in a wide range of wavenumbers $k$. In the linear regime, the surface profile reads:
\begin{equation}
\hat Z(k,t)=\hat Z(k,t=0) e^{\sigma(k)t}
\end{equation}
If the distribution of initial amplitude is initially sufficiently flat and if the growth rate $\sigma(k)$ presents a sharp {\it absolute} maximum in $k_{\rm max}$, then a pattern dominated by the corresponding wavelength $\lambda_{\rm max}$ emerges, as this mode grows the fastest. In the present case, the amplitude of this secondary maximum close to the resonance is almost the same as the value of $\sigma$ at the same wavenumber in the unbounded case. Moreover, most of the modes between the resonance and the ripple peak are in fact much more unstable: if a linear instability could be invoked, the amplitude of all these intermediate modes would eventually be larger than that of this local maximum. Furthermore, the ratio of the primary and the secondary maxima of the growth rate is on the order of $(H/L_{\rm sat})^2$ which is a large number: it is typically on the order of $10^{4}$ for flume or small river experiments ($d\simeq400~\mu$m and $H\simeq40$~cm) and $10^{6}$ for a large river ($d\simeq400~\mu$m and $H\simeq4$~m). For example, taking half a minute for the characteristic ripple apparition time (see Fig.~\ref{AmplitudeTime}a), it would give $\sim 3$ days for the dune linear growth time-scale, i.e. much too large in comparaison to observations (see Fig.~\ref{WavelengthTime}). For these reasons, this secondary peak in the dispersion relation cannot be associated to dunes.

Let us contrast this subaqueous situation to that of aeolian ripples superimposed on aeolian dunes. As already mentioned, the instability mechanism of these dunes is of the same hydrodynamics nature as that of subaqueous ripples, i.e. comes from the upwind shift of the basal shear stress with respect to relief. Aeolian ripples, however, are generated by a screening instability: the upwind face of ripples receives more impacts of saltating grains than the downwind face (\cite{B41,A87,An90}). Rapidely, non-linearities makes the ripple pattern saturate to a wavelength much smaller than that at which dunes emerge, and this saturation is faster than the dune time formation. One can then consider that dunes results from the linear instability of a flat bed which presents saturated aeolian ripples.
Moreover, the growth rate at the wavelength of ripples in the dune instability is negative, because ripples are smaller than the aeolian saturation length. Conversely, the growth rate at the wavelength of dunes in the ripple instability is much smaller than the growth rate at the same wavelength in the dune instability. In conclusion, aeolian ripples and aeolian dunes can truly be associated to two different linear instabilities. In the case of subaqueous ripples and dunes, none of these criterions (different destabilizing mechanisms, saturation of ripples wavelength, separation of the modes by several decades of almost non-growing wavelengths) is full-filled.

%________________________________________________________________________
\section{Field experiments}

In this last section, we present direct experimental evidences that river dunes do not form by a linear instability. We will discuss our field measurements in the light of the model proposed here and show reciprocally that all existing observations are consistent with this model. We will finally propose a new definition of the different subaqueous bedforms, based on the physical mechanisms which control their formation.

%__________________________________
\subsection{Formation of ripples}
We have studied the formation of ripples, dunes and mega-dunes in the Leyre river. This river is located in the south west of France, in a region called `les Landes de Gascogne' ($44^\circ32'$N, $0^\circ52'$W). It flows in a particularly homogeneous basin, both in terms of the nature of the ground (rather sorted sand grains) and of the vegetation. Except during flooding events, bed-load is the dominant mode of transport. The size of the grains on the river bed is around $d=330~\mu$m. The experiments have been performed at the end of summer (low water period), in two straight and flat portions of the river.

\begin{figure}
\includegraphics{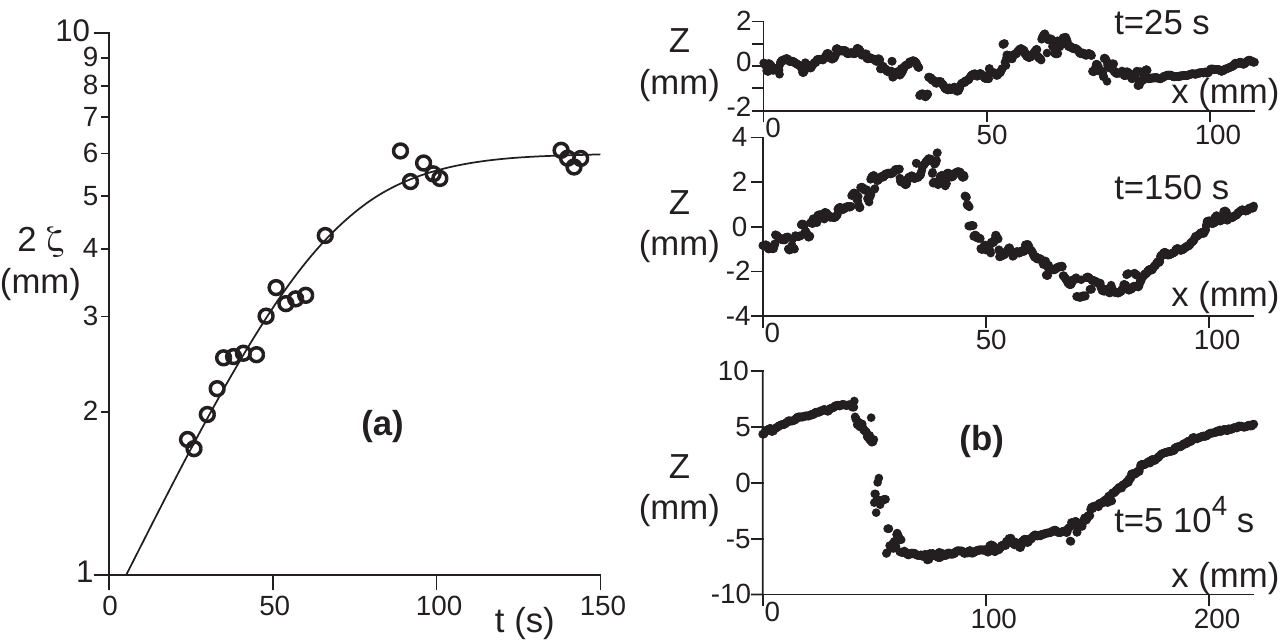}
\caption{Formation of ripples in a natural river, starting at $t=0$ from a flat sand bed. The experiment was performed in the Leyre river, at Sauniac bridge, on September $16^{\rm th}$ 2008,  for $H=52$~cm, $\Fr=0.21$ and $u_*=3$~cm/s. The sediment is well sorted with a mean grain size $d=320\pm70~\mu$m (a) Amplitude $2\zeta$ of the bed disturbances as function of time $t$. $\zeta$ is determined from the auto-correlation of the bed profile $Z(x,t)$. The solid line is the best fit by equation ~(\ref{transient}) and gives $\sigma=3~10^{-2}~{\rm s^{-1}}$ or equivalently $\sigma^{-1}=35~$s. (b) Bed elevation profile measured by taking a picture of the bed enlightened by an inclined laser sheet. Low amplitude ripples can be detected from $t=25$~s after flattening the sand bed. Non-linear effects make the amplitude saturate around $t=100~s$ and a clear avalanche slip face can be observed at $t=150$~s. During this linear stage, the wavelength $\lambda \sim 90$~mm does not evolve significantly (see Fig.~\ref{WavelengthTime}b). Beyond this initial phase, a pattern coarsening toward dunes is observed which saturates at $\lambda\simeq 20$~cm and $\zeta=4.5$~mm after typically one hour: tracking the pattern during further $\sim 14$~hours, we did not observe any significant change of these characteristics.}
\label{AmplitudeTime}
\end{figure}

\begin{figure}
\includegraphics{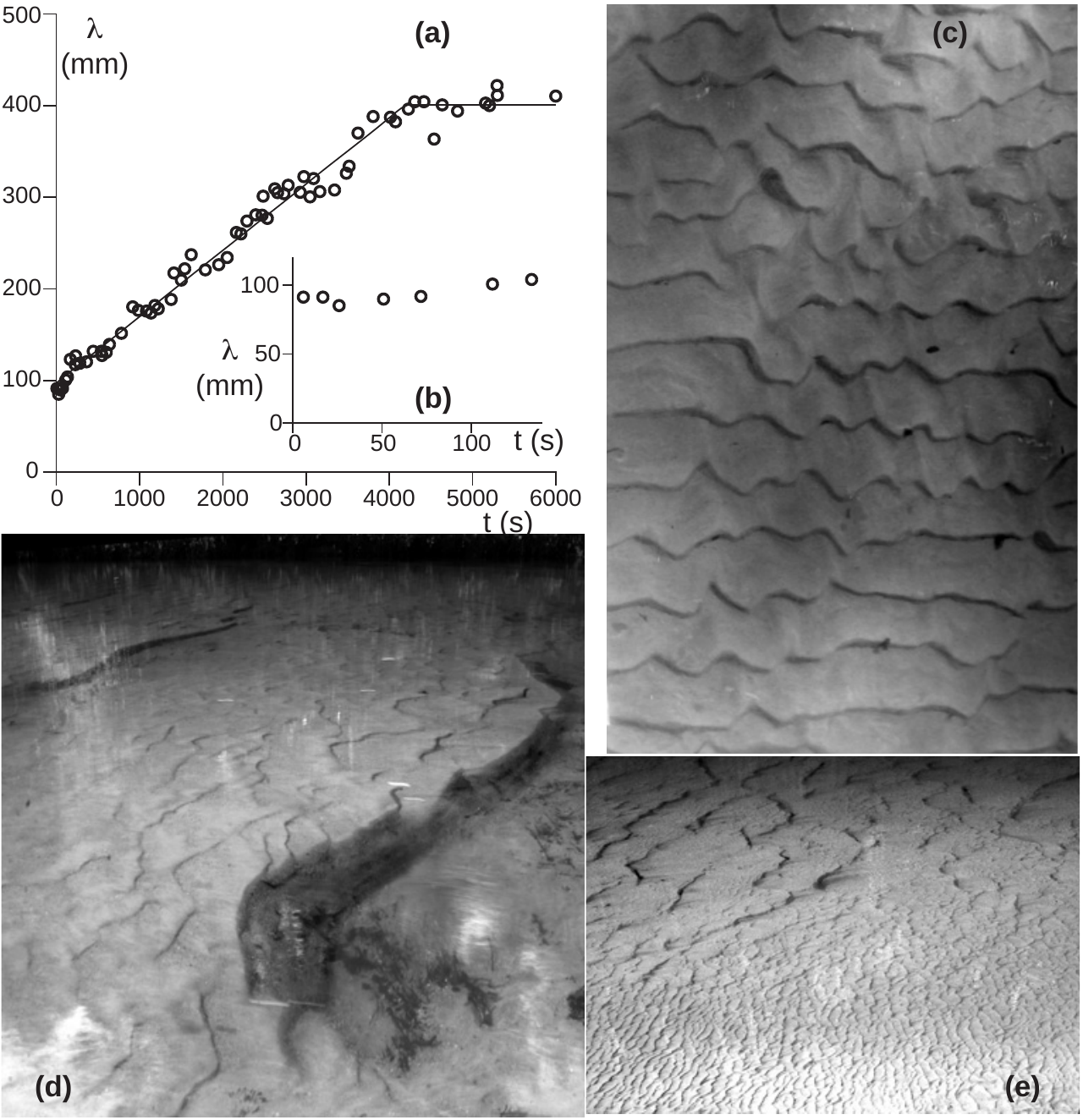}
\caption{(a-c) Formation of dunes in a natural river, starting at $t=0$ from a flat sand bed. The experiment was performed in the Leyre river, at Mios bridge, on September $17^{\rm th}$ 2008, for $H=50$~cm, $\Fr=0.28$ and $u_*=4$~cm/s. The grain size is $d=330\pm70~\mu$m. (a) Time evolution of the wavelength $\lambda$. The pattern coarsening starts after $150~s$ and stops after $\sim 4000$~s. (b) Same graph, but restricted to the linear regime (between $t=0$ and $t=150$~s). (c) The photograph shows the dunes of wavelength $40$~cm formed after $6000$~s. (d) Formation of mega-dunes, starting from a flat sand bed. The experiment was performed in the Leyre river, at Mios bridge, for $H=44$~cm, $\Fr=0.30$ and $u_*=4$~cm/s. The sand is polydisperse: it a mixture of sand grains of size $\sim 330\pm70~\mu$m, which cover $60\%$ of the surface, and of coarse grains larger than $600~\mu$m, which represent $40\%$ of the surface --~but $9\%$ of the grains and $60\%$ of the mass. The photograph shows $3$~m long mega-dunes with $\sim 40$~cm superimposed dunes. (e) Photograph of the Leyre river at Mios bridge showing the sharp transition between dunes (zone of medium sand) and mega-dunes (zone of medium and coarse sand mixed).}
\label{WavelengthTime}
\end{figure}

The experiments were conducted as follows. Using long parallel metallic bars, the surface of the sand bed was carefully leveled at time $t=0$, and the formation of bedforms was directly observed. In order to measure the emergence of ripples, we used a (water proof) laser sheet inclined at a low angle to the horizontal. Taking pictures through a glass plate from the top, we have determined the height profile $Z(x,t)$ along this line as a function of time. Figure~\ref{AmplitudeTime}b shows the evolution of one such a ripple from the initial stage where it is symmetric to the time at which an avalanche slip face develops. To determine the wavelength $\lambda$ and the amplitude $2\zeta$ of the ripple, we computed the auto-correlation of the profile $C(\delta)=\left < Z(x)Z(x+\delta) \right >$. Typically $25$~s after the beginning of the experiment, $C$ shows a secondary maximum whose position gives $\lambda$ and whose amplitude gives $\zeta$. During the first $150$~s, the wavelength does not evolve much whereas the amplitude grows and saturates (Fig.~\ref{WavelengthTime}b). As the final amplitude is not very large compared to the grain size $d$ (and thus the initial noise level), it is difficult to get a convincing evidence of an exponential growth. Still, the curves we obtained are consistent with a linear regime over a factor of two in amplitude (Fig.~\ref{AmplitudeTime}). From our non-linear analysis, we expect an amplitude equation of the form:
\begin{equation}
\frac{d \zeta}{dt}=\sigma \zeta \left[1-\left(\frac{\zeta}{\zeta_\infty}\right)^2\right],
\end{equation}
whose solution is:
\begin{equation}
\zeta = \frac{\zeta_\infty}{\sqrt{1+\exp(-2\sigma t)}} \, .
\label{transient}
\end{equation}
One can see in figure~\ref{AmplitudeTime} that the fit of this relation to the data is very good, so that observations are consistent with a formation of ripples by a linear instability saturated by non-linear hydrodynamical effects. Importantly, the rescaled growth rate $\sigma /(k u_*)$ is around $10^{-3}$ and the rescaled propagation speed $\omega/(k\,u_*)$ around $10^{-2}$. As shown in Part~1, with such small dimensionless numbers, the motion of the bed can be ignored in the hydrodynamical treatment.

%

%__________________________________
\subsection{Formation of dunes and mega-dunes}
We have observed the evolution of the patterns for typically two hours after flattening the sand bed (Fig.~\ref{WavelengthTime}a). A statistically stationary state is eventually reached, which corresponds to what was observed in the natural conditions, i.e. before the experiment (Fig.~\ref{WavelengthTime}c). As in flume experiments (\cite{VCB05a, VCB05b, VL05}), we have observed a coarsening of the ripple pattern, i.e. a progressive growth of the wavelength by merging of bedforms (\cite{RW90,R06}). Figure~\ref{AmplitudeTime}a shows that this growth is linear in time and stops when the wavelength $\lambda$ becomes on the order of the flow depth $H$. Both these processes and the time-scales over which they take place are consistent with the observations of \cite{GSR66} for flume experiments at larger Froude numbers. Again, with a bed motion at the scale of hours, the approximation of a flow over a steady relief is almost perfect. Consistently with the theory, we did not observe the emergence of wavelengths directly at the scale of the flow depth\footnote{In the flume experiments of \cite{CRI05}, the large grain size ($5$~mm) and the moderate water depth ($0.375$~m) are such that the typical size of the first emerging bedforms (few decimeters) are comparable.}. We thus reach the conclusion that the formation of dunes should not be associated to a linear instability but to a non-linear pattern evolution. In the unbounded case (an infinite flow depth), it is probable that this pattern coarsening would have no limit as it is driven by hydrodynamics, which is mostly self-similar. The observation that this coarsening stops at some final wavelength should therefore be associated to a stabilising mechanism, namely the presence of the free surface.

\begin{figure}
\center{\includegraphics{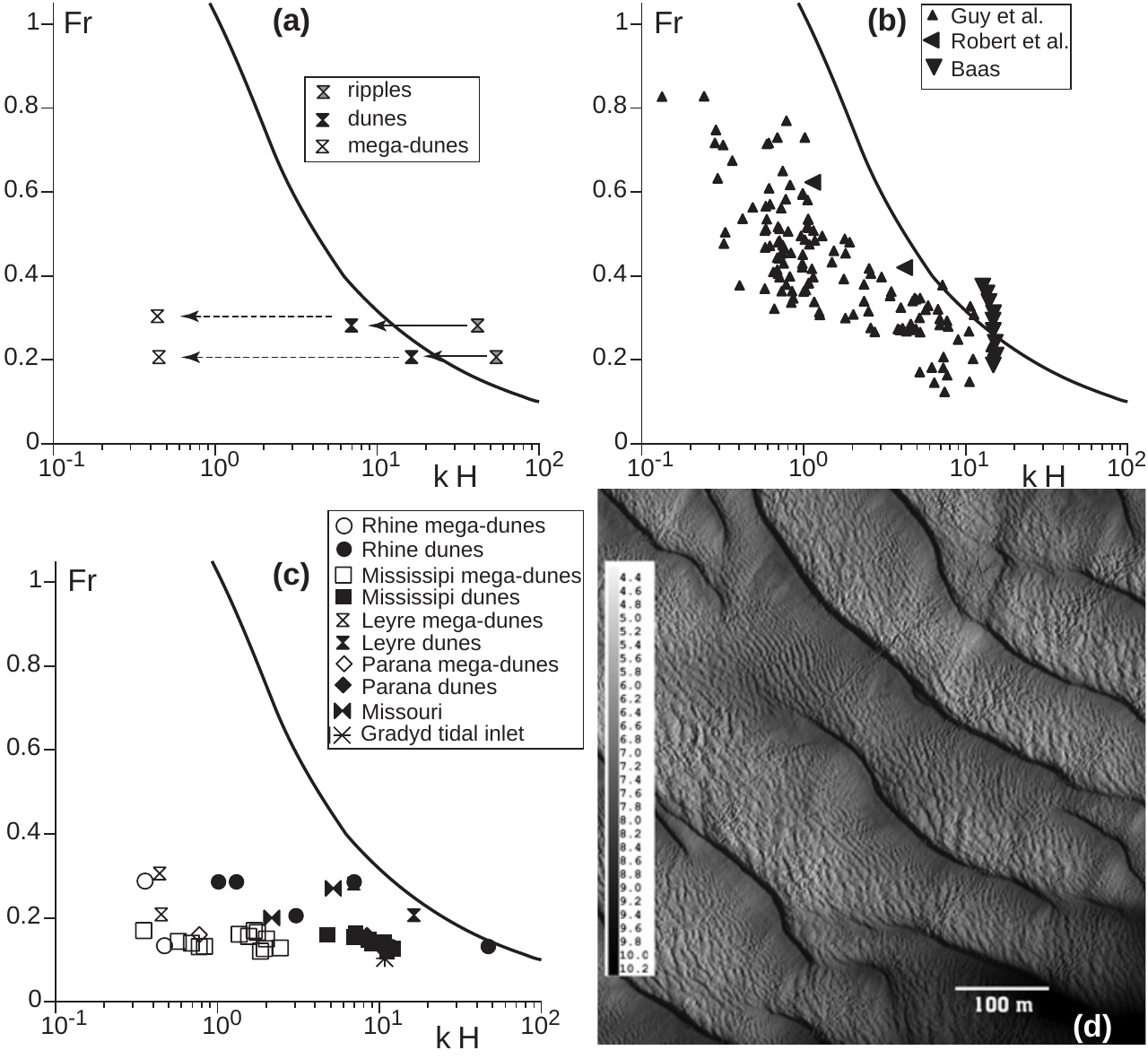}}
\caption{Representation in the $\Fr$ vs $kH$ diagram of different data set. (a) Experiments performed in the Leyre river, starting from a flat sand bed. (b) Flume experiments: wavelength of the stationary bedforms obtained at long time. (c) Bedforms in natural rivers. The mega-dunes are represented by white symbols whereas dunes are displayed with black symbols. (d) Water depth in rio Paran\'a ($H\sim8$~m, $d\sim300~\mu$m, $\Fr=0.16$) measured by multibeam echo sounding (this photograph is from \cite{PBOHKL05}). Mega-dunes of wavelength $\lambda\sim125~{\rm m}\sim 15H$ can be observed, with superimposed dunes of wavelength $6$~m and probably superimposed ripples too (not visible).}
\label{RideDune}
\end{figure}

The places selected for the experiments were of particular interest as different final wavelengths were observed across the river (Fig.~\ref{WavelengthTime}c-e). The experiments just described above have been performed in of the side of the river (say, at a distance less than the third of the river width from the bank). In this `external' region, the sand is well sorted and the dune wavelength is observed to saturate for a rescaled wave-number $kH$ just below the resonant conditions (Fig.~\ref{RideDune}a). By contrast, in the central part of the river, much larger bedforms are present (Fig.~\ref{WavelengthTime}d-e). They display superimposed dunes on their stoss side, and we call mega-dunes here-after. The river slope and the flow velocity were not significantly different in the dune and mega-dune regions. The major difference was the presence of coarse grains causing bed armouring in the central part of the river. As shown in figure~\ref{WavelengthTime}e, the transition between the two regions is rather sharp. We have flattened the bed over a zone of $12$~m in length and $4$~m in width to observe the formation of mega-dunes. The initial stage is the formation of ripples composed of small sand grains that merge, leading to the same dunes as described on the side of the river. However, in the course of this pattern coarsening, the inter-dune zone becomes richer in coarse grains so that the dunes eventually propagate on a bed which is more difficult to erode. They progressively amalgamate into mega-dunes  of wavelength ten times larger than the dunes, covered with superimposed ripples and dunes. Even in the asymptotic state, superimposed bedforms are continuously generated. As they propagate faster than the mega-dune, they accumulate at its crest. During the transient of formation of mega-dunes (typically $5$~hours in our experiments), the pattern is disordered and is not composed by a unique Fourier mode. As far as one can say without having explicitly performed a multi-scale analysis of the topography, structures of growing size were progressively formed, which become ordered as they reach the final mega-dune wavelength (between $10H$ and $20H$).

In summary, a small difference in the experimental conditions (here, most probably, the presence of coarse grains or not) can significantly affect the pattern coarsening dynamics. The wavelength of mega-dunes can be larger than those of dunes by one to two orders of magnitude (Fig.~\ref{RideDune}a). With this observation, the question is not anymore to find a new destabilising mechanism to explain the dune formation. Rather, as larger and larger bedforms are produced by non-linear processes, one needs to identify a stabilising mechanism limiting this coarsening. As evidenced here, this is exactly the role played by the free surface.
%

%__________________________________
\subsection{Conclusion}
Figure~\ref{RideDune} presents several series of measurements of the final bedform wavelength collected in the literature for flume experiments (panel b) and natural rivers (panel c). The most impressive data-set is certainly that reported by~\cite{GSR66}. As these authors separated their data into ripples and dunes on the basis of a definition that we do not agree with (see introduction), we have ignored here their annotations. Importantly, \cite{GSR66} have not looked at the transient of formation of bedforms but have focused on the long time regime. For example, they have often started a new experiment with a bed in the state reached at the end of the previous run. Moreover, the control parameters (slope and flow rate) were varied in the course of the experiment to maintain constant secondary quantities such as the flow depth or the Froude number. This methodology is probably responsible for the lack of reproducibility and the huge dispersion of data (one decade horizontally) on can see in figure~\ref{RideDune}b. By contrast, the series of experiments performed by \cite{B99} are much more controlled and reproducible. In particular, the wavelength is measured as a function of time starting from a flat sand bed and fitted to obtain its asymptotic value. These flume experiments clearly evidence the difference between the initial wavelength at which ripples form, their evolution by pattern coarsening to form dunes and the non-linear selection of the final wavelength due to the free surface. Note that we have not taken into account the correction of raw measurements performed by \cite{B99} to take into account the temperature dependence of the viscosity, as it is a non-sense in the hydraulically rough regime. The last points in figure~\ref{RideDune}b have been obtained by \cite{RU01}. As for  \cite{GSR66}, we have not taken into account the denomination of the bedforms (ripple or dunes) used in this article. The points obtained by \cite{B99} and \cite{RU01} are very close to the resonant conditions; the whole data set of \cite{GSR66} is clearly in the subcritical regime --~at low $\Fr$ and/or low wavenumber $kH$~-- and globally follows the resonance curve. We have gathered in figure~\ref{RideDune}c the points measured in the Leyre river, in the Gr\aa dyd tidal chanel (\cite{Ba05}), in the Mississipi river (\cite{H98}), in the Missouri river (\cite{A72}), in the Rhine (\cite{CGOR00,WB03}) and in the Rio Paran\'a (\cite{PBOHKL05}). One can observe that the dunes propagating on the stoss slope of mega-dunes lie in the same region of the diagram as the simple dunes --~roughly between $kH=0.1/\Fr^2$ and $kH=1/\Fr^2$. The mega-dunes lie between $kH=0.3$ and $kH=0.1/\Fr^2$. It should be emphasised that the different points in this graph correspond to very different flow depth $H$ (compare e.g. the megadunes in the Leyre river shown in figure~\ref{AmplitudeTime}d and those in the rio Paran\'a shown in figure~\ref{RideDune}d). It means that the saturation length, which determines the wavelength at which ripples form, is not a relevant length for the formation of dunes and mega-dunes.
\begin{figure}
\center{\includegraphics{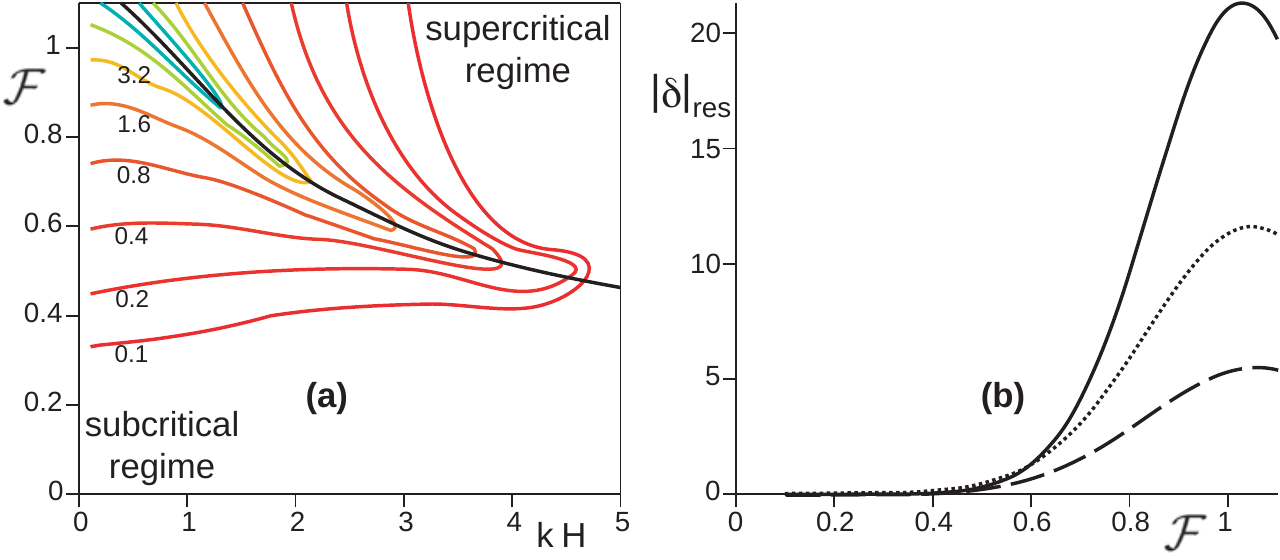}}
\caption{Amplitude  $|\delta|$ of the distortion of the free surface related to that of the bedform. (a) Isocontours of $|\delta|$ in the plane $\Fr$ vs $kH$ (color online). (b) Distortion $|\delta|$ computed along the resonance curve (black solid line in panel a) for $H/z_0=10^2$ (dashed line), $H/z_0=10^3$ (dotted line) and $H/z_0=10^4$ (solid line) as a function of the Froude number $\Fr$. The amplitude in the resonant conditions is maximum at $\Fr\simeq1$. }
\label{ResonanceFrkH}
\end{figure}

We then see why a distinction between ripples and dunes based on some absolute dimension considerations are misleading from the physical point of view: depending on the value of the water depth, bedforms `of less than about $2$ feet' (\cite{GSR66}), or `less than $0.6$~m' (\cite{A90}), can feel or not the free surface. Nor a criterion based on the amplitude $|\delta| \zeta$ of distortion of the free surface can be satisfying to define dunes. Figure~\ref{ResonanceFrkH} shows the value of $|\delta|$ in the parameter space ($\Fr$,$kH$), computed using the hydrodynamical model detailed in part 1. One observes that the free surface is distorted by less than $10\%$ of the bedform amplitude for $\Fr<0.35$ or for $kH>3$. With a definition based on the amplitude of distortion of the free surface such as that proposed by \cite{GSR66}, there would be no dune at all in natural rivers! At low Froude numbers, dunes do not distort significantly the free surface. Approaching $\Fr=1$ from below, the distortion of the free surface, in antiphase with the topography, becomes more and more pronounced (Fig.~\ref{ResonanceFrkH}b). Above $\Fr\simeq0.6$, this effect becomes sufficient to create a zone of stable wavelengths around the resonance (Fig.~\ref{FrkH}) and thus a gap to cross during the pattern coarsening.

These results then suggest a new classification of river bedforms based on the dynamical mechanisms responsible for their formation. The most obvious criterion is the sensitivity to the presence of the free surface. We thus define ripples as the bedforms whose wavelength $\lambda$ is sufficiently small compared to the flow depth $H$ not to feel the finiteness of the flow depth. In the diagram $\Fr$ vs $kH$, they are located in the supercritical region, i.e. their wavenumber is larger than the resonant value. When the scale separation between the flow depth and the saturation length is sufficient, the structures that form by linear instability of a flat sand bed are ripples. Beyond the linear regime the pattern coarsening leads to growing wavelengths that reach the resonance curve. Once in the subcritical region on the left of this curve (see Fig.~\ref{RideDune}) hydrodynamics becomes affected and in some case dominated by the presence of the free surface and the bedforms can be called dunes. As evidenced by our field experiments, the pattern coarsening can end with very different bedforms depending on the conditions (e.g. the grain size distribution). The associated non-linear selection of pattern wavelength is an open problem (see \cite{PM04,ACP06}) that will require specific investigations. Our observations suggest to define mega-dunes as bedforms sufficiently large to present superimposed dunes. With this definition, what is usually called `river bars' in the literature falls into our category of mega-dunes. Similarly, the term `sand-wavelet' introduced by Coleman (see e.g. \cite{CM94,CM96}) are ripples in their initial stage. It is furthermore worth noting that dunes can present superimposed ripples when the scale separation between the flow depth $H$ and the saturation length $L_{\rm sat}$ is sufficient. The criterion to separate dunes from mega-dunes is then based on the location of the superimposed bedforms with respect to the resonance curve in the plane $\Fr$ vs $kH$. This distinction is important as in very deep water, the pattern coarsening would lead to ripples large enough to accommodate superimposed ripples on their stoss side. In this case, one should talk about mega-ripples. In summary, the appellation of given subaqueous bedforms should be chosen in function of their location in the diagram $\Fr$ vs $kH$, with a suffix 'mega-' to point out the presence or not of superimposed structures.

%________________________________________________________________________
\section{Summary of the results}

As the present paper aims both to present novel results and to discuss the state of the art of the physics of ripples and dunes, we wish to sum up in this section, in a qualitative manner, our methodology and key results.

We have shown that the linear instability of a flat sand bed can be abstracted in terms of four well-defined quantities:
\begin{itemize}
\item the component of the basal shear stress in phase and in quadrature with the bottom (the parameters $A$ and $B$, see part 1),
\item the relation between the sediment flux, the shear stress and its threshold value, which determines the time-scale of the instability but does not influence the initial wavelength,
\item the saturation length, i.e. the relaxation length between the sediment flux and the shear stress, which essentially determines the length-scale of the instability,
\item the relation between the threshold shear stress and the slope, which encodes the stabilising effect of gravity.
\end{itemize}
The framework resulting from these ingredients is very general and can encompass a wide spectrum of specific models without any profound difference of nature. As discussed in the first part of these twin papers, the details of the hydrodynamical description can affect $A$ and $B$ to various degrees, depending for instance on the physics at work in the surface layer or on the way the different length-scales of the problem intermix with each other -- see the summary section of part~1. From the choice of a particular transport model, one can compute the steady (or saturated) value of the sediment flux, as well as the characteristic length over which the actual flux relaxes to equilibrium. Finally, the expression of the threshold shear stress as a function of the grain size, the bed slope and other possible parameters (e.g. the cohesion between the grains) comes from a separate mechanical model. Doing so, one can then independently investigate the effect of each dynamical mechanism on the formation of ripples and dunes.

We have devoted a large fraction of this part~2 to the study of two sediment transport regimes. They differ by the nature of the mechanisms that make the flux saturate:
\begin{itemize}
\item an erosion limited regime, in which the negative feedback of the particle transport on the flow is negligible. The saturation of the flux is limited by the time needed to entrain one grain at the surface of the sand bed (\cite{C06}). The associated saturation length is determined by the distribution of the traps at the surface of the bed. It is proportional to the grain size and it should diverge when the shear velocity reaches the value for which the flow is sufficient to entrain all the grains at the surface.
\item a momentum limited regime, in which the flux saturates due to the extra-shear stress exerted by the moving grains on the fluid (\cite{B56}). If the shear velocity is not too large so that bed load is the dominant mode of transport, the saturation length is dominated by the grain inertia thus proportional to the density ratio between the grain and the fluid times the grain size.
\end{itemize}
We have derived a detailed model of bed load describing both the erosion limited and the momentum limited regimes that is able to fit the data of \cite{J98} and \cite{LvB76}. We have also derived a quantitative model for the threshold Shields number, which takes into account the bed slope effect and is able to reproduce the observed decrease at the transition between the viscous and the turbulent regimes. Sheet flows, in which the grains flow over a significant number of grain diameters, and suspended load are outside the validity of the model because they require a different hydrodynamical treatement where the flow inside the transport layer must be modelled.

These different ingredients at hand, we have performed the linear stability analysis of a flat sand bed. The destabilising mechanism is of hydrodynamical nature and is related to the phase advance of the basal shear stress with respect to the topography. Two stabilising mechanisms are identified: the sediment transport saturation length $L_{\rm sat}$ and the slope effect, which depends on the ratio $u_*/u_{\rm th}$. As $L_{\rm sat}$ is generically larger than $z_0$, it dominates and essentially determines the scaling law followed by the most growing wavelength $\lambda_{\rm max}$. This most unstable mode is associated to ripples, which thus form by a linear instability. In the case of a smooth bottom, the roughness seen from the inner layer is governed by the viscous length $\nu/u_*$, which may dominate the scaling of $\lambda_{\rm max}$ if $\nu/u_*$ is larger than $L_{\rm sat}$ (\cite{SB84}).

Due to the slope effect, the ratio $\lambda_{\rm max}/L_{\rm sat}$ is a decreasing function of $u_*/u_{\rm th}$. We have analysed different sets of measurements of initial ripple wavelengths $\lambda_{\rm max}$ available in the literature and deduced the corresponding saturation length, assuming that the hydrodynamical model is correct. $L_{\rm sat}$ is found roughly independent of $u_*$ and between $5~d$ and $15~d$. These values are consistent with a saturation length limited by the grain inertia ($L_{\rm sat}\simeq 2 (\rho_s/\rho_f)\,d$) as previously stated by the authors (\cite{CA06,AC07}). This prediction is especially good for large grains ($d \gtrsim 1~$mm), for which viscous effects are completely negligible. However, the data show systematic dependencies that are not captured by the present model: the wavelength at which ripples form is systematically larger for smaller or rougher grains.

We have performed field experiments in the Leyre river, whose results show that the evolution of the ripple wavelength $\lambda$ and amplitude $\zeta$ at short time are consistent with a linear instability. We also observed that these ripples present a pattern coarsening: their wavelength grows and saturate just after crossing the resonance condition of surface waves. In the course of this pattern coarsening, the bedforms are in quasi-equilibrium between erosion and deposition, a situation for which we have shown that the aspect ratio of ripples can be predicted quantitatively using the weakly non-linear description of hydrodynamics derived in part~1.

The influence of the river free surface on the bed is stabilising. As shown in part~1 and in agreement with standard hydraulics, the gravity waves excited at the free surface by the bedforms are in phase at small $\lambda/H$ (supercritical regime) and in antiphase at large $\lambda/H$ (subcritical regime). In between, the free surface is phase-advanced with respect to the bottom, so that it tends to induce a phase-delay of the shear stress on the ground. Bedforms of wavelength around the resonance conditions are thus stabilised. Moreover, at very large $\lambda/H$, the inner layer invades the whole flow and the free surface again has a strong stabilising effect. As no destabilising mechanism is associated to the presence of the free surface, dunes do not form by a linear instability. This result is directly confirmed by our field experiments showing that they form by non-linear pattern coarsening, as suggested by \cite{RW90} and \cite{R06}.  Finally, our experiments show that the non-linear selection of the final wavelength is very sensitive to small changes in the experimental conditions, and in particular to the presence of coarse grains.

Finally, our results suggest to classify subaqueous bedforms according to the dynamical mechanisms that control their formation. We thus propose the following bedform definitions and characteristics:
\begin{itemize}
\item Ripples are bedforms whose wavelength $\lambda$ is sufficiently small compared to the flow depth $H$ not to feel the presence of the free surface. In the diagram $\Fr$ vs $kH$, they are located on the right of the resonance curve (supercritical regime). They form by linear instability and their initial wavelength essentially scales on the saturation length $L_{\rm sat}$. They exhibit pattern coarsening and remain ripples until they cross the condition of resonance of the surface waves.
\item Dunes are bedforms whose wavelength $\lambda$ is sufficiently large compared to the flow depth $H$ to be stabilised by the presence of the free surface. The ripple pattern coarsening generically leads to the formation of dunes. In the diagram $\Fr$ vs $kH$, they are located along the resonance curve on the subcritical side. If the flow depth $H$ is much larger than the wavelength at which ripples form, dunes may present superimposed ripples.
\item Like dunes, mega-dunes are under the influence of the free surface and on the left side of the resonance curve but they present superimposed dunes. They typically result from the coarsening of a dune pattern pushed to very large wavelength by heterogeneities (in particular a polydispersed sediment). 
\end{itemize}

Further studies are needed to investigate this non-linear wavelength selection in details. In the purpose of describing the interactions between bedforms, future models will have to incorporate hydrodynamical non-linearities and in particular flow separation, for which the systematic expansion with respect to the amplitude performed in part~1 may serve as a well-controlled starting point.

\vspace*{0.3cm}

%____________________________________________________________________________
\noindent
\rule[0.1cm]{3cm}{1pt}

The field work has been carried out in agreement with the Parc Naturel R\'egional des Landes de Gascogne. This work has benefited from the financial support of the french minister of research.

\appendix

%________________________________________________________________________
\section{Static threshold}
\label{Staticthreshold}

In this appendix, we derive an explicit expression of the transport threshold, valid both in the viscous and turbulent regimes. We introduce the effective velocity $U$ of the flow around the sand grains and define for convenience its normalised counterpart:
\begin{equation}
\mathcal{U} = \frac{U}{\sqrt{(\rho_s/\rho_f-1) gd}} \, .
\label{Uadim}
\end{equation}
For the sake of simplicity, we will take for $U$ the fluid velocity at the altitude $z=d/2$. A good approximation of the relation between the shear velocity $u_*$ and the typical velocity around the grain $U$ is given by
\begin{equation}
u_*^2=\frac{2 \nu }{d}  U+ \frac{\kappa^2}{\ln^2(1+1/2r)} U^2 \, .
\label{relationshearU}
\end{equation}
where $\nu$ is the kinematic viscosity of the fluid, $\kappa$ the von K\'arm\'an constant and $r$ the aerodynamic roughness rescaled by the grain diameter $d$. The rescaled bed roughness $r$ is on the order of $r=1/10$.

At the threshold, the horizontal force balance on a grain of the bed reads
\begin{equation}
\frac{\pi}{6}  \mu ( \rho_s - \rho_f ) g d^3= \frac{\pi}{8}  C_d  \rho_f U_{\rm th}^2 d^2,
\end{equation}
where $\mu$ is the avalanche slope for sand grains, i.e. $\tan32^\circ$. $C_d$ is the drag coefficient, which is a function of the grain Reynolds number $\mathcal{R}$. With a good accuracy, the drag law for natural grains can be put under the form:
\begin{equation}
C_d = \left(C_\infty^{1/2}+s \mathcal{R}^{-1/2} \right)^2 \quad{\rm with \quad} \mathcal{R}= \frac{U d}{\nu}\, .
\label{dragcoeff}
\end{equation}

At this stage, we introduce the viscous size $d_\nu$, defined as the diameter of grains whose
free fall Reynolds number $u_{\rm fall}d/\nu$ is unity:
\begin{equation}
d_\nu=(\rho_s/\rho_f-1)^{-1/3}~\nu^{2/3}~g^{-1/3}.
\end{equation}
It corresponds to a grain size at which viscous and gravity effects are of the same order of magnitude. From the three previous relations, we get the equation for $\mathcal{U}_{\rm th}$:
\begin{equation}
C_\infty^{1/2} \mathcal{U}_{\rm th} +s  \left(\frac{d_\nu }{d}\right)^{3/4}  \mathcal{U}_{\rm th}^{1/2} = \left(\frac{4\mu}{3}\right)^{1/2},
\end{equation}
which solves into
\begin{equation}
\mathcal{U}_{\rm th} =\frac{1}{4 C_\infty}\left[\left( s^2\left(\frac{d_\nu }{d}\right)^{3/2} +8 \left(\frac{\mu C_\infty}{3}\right)^{1/2}\right)^{1/2}-s  \left(\frac{d_\nu }{d}\right)^{3/4}\right]^2.
\label{uth1}
\end{equation}
The corresponding expression of the static threshold Shields number is finally:
\begin{equation}
\Theta_{\rm th} =2 \left(\frac{d_\nu }{d}\right)^{3/2}  \mathcal{U}_{\rm th} +  \frac{\kappa^2}{\ln^2(1+1/2r)} \mathcal{U}_{\rm th}^2.
\label{uth2}
\end{equation}
In the viscous regime, the above relations simplify into:
\begin{equation}
\mathcal{U}_{\rm th}=\left(\frac{4 \mu}{3 s^2} \right) \left(\frac{d}{d_\nu}\right)^{3/2} \quad {\rm and} \quad \Theta_{\rm th} =\frac{8 \mu}{3 s^2} \, .
\end{equation}
In the turbulent regime,  we get:
\begin{equation}
\mathcal{U}_{\rm th}= \sqrt{\frac{4 \mu }{3 C_\infty}} \quad {\rm and} \quad \Theta_{\rm th} = \frac{4 \mu \kappa^2}{3 C_\infty \ln^2(1+1/2r)} \, .
\end{equation}
As can be seen in figure~\ref{Threshold}, the most interesting range of grain diameters, between $100~\mu$m and $1~$mm, is precisely the zone of transition between the two regimes. The crossover grain diameter $d_{co}$ scales as:
\begin{equation}
d_{co}=\frac{s^{4/3}}{(\mu C_\infty)^{1/3}}\,d_\nu,
\end{equation}
which is around $200~\mu$m.

%________________________________________________________________________
\section{Motion of a single grain}
\label{grainMotion}

\begin{figure}
\includegraphics{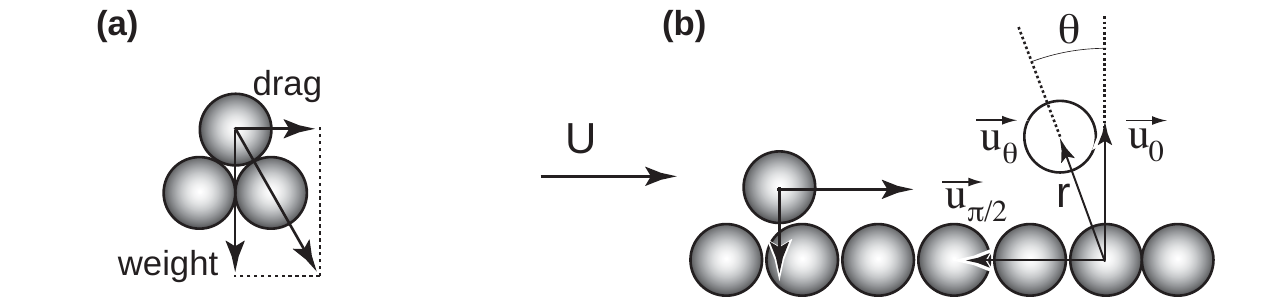}
\caption{Geometry of the trajectory calculation.}
\label{Trajectory}
\end{figure}

To describe the elementary jump by one grain diameter, we write the equation of motion of one grain dragged by a flow of effective velocity $U$ and submitted to gravity. We assume that it looses completely its energy during collisions, due to the thin layer of fluid between the grains. Contrarily to \cite{B56} and \cite{CLDZ08}, we do not introduce any `effective friction': instead, we take into account the geometry of the trapping grains (Fig.~\ref{Trajectory}), which is the physical effect responsible for this friction, as for the transport threshold.

Starting from the position of the grain $d \vec u_\theta$, we get its velocity $\vec V = d \dot \theta \vec u_{\theta+\pi/2}$ and its tangential acceleration $d \ddot \theta \vec u_{\theta+\pi/2}$. We assume that the reaction of the substrate is normal to it. Using, as in the appendix~\ref{Staticthreshold},  $d$ as a unit size and $[(\rho_s/\rho_f-1) gd]^{1/2}$ as a unit velocity (see Eq.~\ref{Uadim}), the dynamical equation reads:
\begin{eqnarray}
\ddot \theta &=& \frac{3}{4}  \left[C_\infty^{1/2}  \left(\dot \theta^2-2 \cos \theta\,\mathcal{U}\,\dot \theta +\mathcal{U} ^2\right)^{1/4}+ s  \left(\frac{d_\nu }{d}\right)^{3/4}\right]^2  \left(\cos \theta\,\mathcal{U} -\dot \theta\right)  + \sin \theta \label{EqDyn}\\
&=&   \left[\frac{\sqrt{3}}{2}C_\infty^{1/2}  \left[\left(\dot \theta^2-2 \cos \theta\,\mathcal{U}\,\dot \theta +\mathcal{U}^2\right)^{1/4}-\mathcal{U}_{\rm th}^{1/2}\right]+  3^{-1/4} \mathcal{U} _{\rm th}^{-1/2} \right]^2  \left(\cos \theta\,\mathcal{U} -\dot \theta\right)  + \sin \theta, \nonumber
\end{eqnarray}
where $\mu$ has been taken equal to $\tan(\pi/6)=1/\sqrt{3}$. As in the appendix \ref{Staticthreshold}, the rescaled velocity $\mathcal{U}$ is related to the Shields number by:
\begin{equation}
\Theta =2 \left(\frac{d_\nu }{d}\right)^{3/2}  \mathcal{U} +  \frac{\kappa^2}{\ln^2(1+1/2r)} \mathcal{U}^2.
\end{equation}
This equation can be integrated numerically between $\theta=-\pi/6$ and $\theta=\pi/6$, starting from $\dot \theta=0$. Note that there is no adjustable parameter in this description.

In order to get the asymptotic behavior analytically, we perform an expansion with respect to $\theta$ around $-\pi/6$ and with respect to $\mathcal{U}$ around $\mathcal{U}_{\rm th}$. We get a linear equation of the form:
\begin{equation}
\ddot \theta + \left(\frac{1}{\sqrt{3}\,\mathcal{U}_{\rm th}}+\frac{3^{5/4}}{8} C_\infty^{1/2}\right)  \dot \theta - \frac{2}{\sqrt{3}} \left(\theta+\frac{\pi}{6}\right)= \frac{2\pi \left(\mathcal{U} - \mathcal{U}_{\rm th} \right)}{3^{3/2} \alpha \mathcal{U}_{\rm th}} \, , \nonumber
\end{equation}
with
$$\alpha=\frac{8 \pi }{3^{3/2} \left(2+3^{3/4} C_\infty^{1/2} \mathcal{U}_{\rm th} \right)} \, .$$
The solution of this linear equation is:
\begin{equation}
\theta=-\frac{\pi}{6} + \frac{\pi \left(\mathcal{U}_{\rm th}-\mathcal{U}\right)}{3 \alpha \mathcal{U}_{\rm th}} \left[1+\frac{r_- \exp(r_+t)-r_+\exp(r_- t)}{r_+-r_-}\right], \nonumber
\end{equation}
where $r_-$ and $r_+$ are the solutions of the equation:
\begin{equation}
r^2 +  \left(\frac{1}{\sqrt{3} \mathcal{U}_{\rm th} }+\frac{3^{5/4}}{8} C_\infty^{1/2}\right)  r - \frac{2}{\sqrt{3}}=0. \nonumber
\end{equation}
The time $T$ needed to pass over $d$ thus scales around the threshold as:
\begin{equation}
T \sim \frac{1}{r_+}\,\ln \left[ \left(1-\frac{r_+}{r_-}\right) \left(1+\frac{\alpha \mathcal{U}_{\rm th}}{\mathcal{U}-\mathcal{U}_{\rm th}} \right) \right]\,\sqrt{\frac{\rho_f d}{(\rho_s-\rho_f) g}} \, .
\label{ExpT}
\end{equation}
Making use of equation~(\ref{relationshearU}), one sees that this erosion time diverges at the threshold as:
\begin{equation}
T \propto\,-\ln \left[\Theta-\Theta_{\rm th} \right].
\end{equation}
On the other hand, $T$ should resume to $d/U$ far from the threshold. Integrating numerically equation~(\ref{EqDyn}), we have found convenient approximations that follow the correct asymptotic behaviors just above the threshold and far from it. In the turbulent regime, it reads:
\begin{equation}
T \propto\,\ln \left[\frac{\sqrt{\Theta}+\sqrt{\Theta_{\rm th}}}{\sqrt{\Theta}-\sqrt{\Theta_{\rm th}}} \right]\,\sqrt{\frac{\rho_f d}{(\rho_s-\rho_f) g}} \, ,
\label{ApproxTturb_appendix}
\end{equation}
where the prefactor is around $1.5$. In the viscous regime, it reads
\begin{equation}
T \propto\,\ln \left[\frac{\Theta+\Theta_{\rm th}}{\Theta-\Theta_{\rm th}} \right]\, \left(\frac{d_\nu }{d}\right)^{3/2}  \,\sqrt{\frac{\rho_f d}{(\rho_s-\rho_f) g}} \, ,
\label{ApproxTvisc_appendix}
\end{equation}
with a prefactor around $17$.

%________________________________________________________________________
\section{Saturation length in a continuum description of momentum limited bed load}

We start from the momentum balance equation, as written by \cite{P75}:
\begin{equation}
\rho_s\left(\partial_t q+\frac{q}{h_0} \partial_x q\right)=\rho_f u_*^2-\tau_{\rm sand}(q).
\end{equation}
By definition, the flux $q$ saturates to $q_{\rm sat}$ when the shear stress is equal at the top of the transport layer and below it, on the bottom:
\begin{equation}
\tau_{\rm sand}(q_{\rm sat}(u_*))=\rho_f u_*^2.
\end{equation}
Linearising the dynamical equation, we get:
\begin{equation}
\rho_s\left(\partial_t q+\frac{q_{\rm sat}}{h_0} \partial_x q\right)=\left.\frac{d\tau_{\rm sand}}{dq}\right|_{q=q_{\rm sat}}\,(q_{\rm sat}-q).
\end{equation}
Identifying this equation with the definition of the saturation length $L_{\rm sat}$ and the saturation time $T_{\rm sat}$,
\begin{equation}
T_{\rm sat} \partial_t q+L_{\rm sat} \dr_x q = q_{\rm sat} - q,
\label{charge_appendix}
\end{equation}
we get:
\begin{equation}
L_{\rm sat} =\rho_s \frac{q_{\rm sat}}{h_0}\,\frac{dq_{\rm sat}}{d\tau}=\frac{\rho_s}{\rho_f} \frac{q_{\rm sat}}{2\,h_0\,u_*}\,\frac{dq_{\rm sat}}{d u_*} \, .
\end{equation}
One readily see that $L_{\rm sat}$, as predicted by this expression, vanishes at the transport threshold as $q_{\rm sat}$ and increases very fast with $u_*$. None of these behaviors is correct, whatever the shape of the bed load equation $q_{\rm sat}(u_*)$ --~Einstein-Brown, Bagnold or those derived in the text. There are two simple reasons for this failure. First, the continuum approach ignores the process by which the momentum is transferred from the fluid to the grains (namely the drag of the grains). Second, the flux $q$ evolves both because the number of moving particles evolves and because their velocity changes. A valid description should thus both present an equation for the erosion and a description of the trajectories, as done in the text.

%\newpage

%________________________________________________________________________

\end{document}